\documentclass[final,5p,times,twocolumn]{elsarticle}

\usepackage{hyperref}
\usepackage{subfigure}
\usepackage{graphicx}
\usepackage{amsmath,amssymb}
\usepackage{siunitx}
\usepackage{cleveref}
\usepackage{tikz}
\usetikzlibrary{patterns}
\usetikzlibrary{plotmarks}
\usepackage{color}
\usepackage{ulem}
\usepackage{comment}
\usepackage{bm}

\usepackage[version=4]{mhchem}

\usepackage[utf8]{inputenc}
\usepackage{pgfplots}
\DeclareUnicodeCharacter{2212}{−}
\usepgfplotslibrary{groupplots,dateplot}
\usetikzlibrary{patterns,shapes.arrows}
\pgfplotsset{compat=newest}

\usetikzlibrary{patterns,plotmarks}

\journal{Journal of \LaTeX\ Templates}

\begin{document}

\begin{frontmatter}

\title{Development and Quality Control of PMT Modules for the Large-Sized Telescopes of the Cherenkov Telescope Array Observatory}

\cortext[corresponding]{Corresponding authors: T. Saito ({\tt tsaito@icrr.u-tokyo.ac.jp}), M. Takahashi ({\tt takahashi.mitsunari@isee.nagoya-u.ac.jp}), Y. Inome ({\tt sinome@icrr.u-tokyo.ac.jp})}
\author[ICRR]{T.~Saito\texorpdfstring{\corref{corresponding}}}
\author[ISEE]{M.~Takahashi\texorpdfstring{\corref{corresponding}}}
\author[ICRR]{Y.~Inome\texorpdfstring{\corref{corresponding}}}
\author[ICRR]{H.~Abe}
\author[IFAE]{M.~Artero}
\author[IFAE]{O.~Blanch}
\author[IAC]{J.~Becerra Gonz\'alez}
\author[ICRR]{S.~Fukami\fnref{nowatdesy}}
\author[ICRR]{D.~Hadasch}
\author[ICRR]{Y.~Hanabata}
\author[Ibaraki]{Y.~Hattori}
\author[IAC]{J.~Herrera Llorente}
\author[ICRR]{K.~Ishio\fnref{nowatpo}}
\author[Kyoto]{H.~Iwasaki}
\author[Ibaraki]{H.~Katagiri}
\author[Konan]{K.~Kawamura}
\author[IFAE,cnrs]{D.~Kerszberg}
\author[Tokai]{S.~Kimura}
\author[Saitama]{T.~Kiyomoto}
\author[ICRR]{T.~Kojima}
\author[Kyoto]{Y.~Konno}
\author[ICRR]{Y.~Kobayashi\fnref{nowatchiba}}
\author[Saitama]{S.~Koyama}
\author[ICRR]{H.~Kubo}
\author[Tokai]{J.~Kushida}
\author[IAC]{A.~L\'{o}pez-Oramas}
\author[Kyoto]{S.~Masuda}
\author[Saitama]{S.~Matsuoka}
\author[ICRR]{D.~Mazin}
\author[ICRR]{D.~Nakajima}
\author[Yamagata]{T.~Nakamori}
\author[Saitama]{T.~Nagayoshi}
\author[IFAE]{D.~Ninci}
\author[Tokai]{K.~Nishijima}
\author[Saitama]{G.~Nishiyama}
\author[Ibaraki]{Y.~Nogami}
\author[MPI]{S.~Nozaki}
\author[ICRR]{M.~Ogino}
\author[ICRR]{H.~Ohoka}
\author[Kyoto]{T.~Oka\fnref{nowatritsu}}
\author[Ibaraki]{S.~Ono}
\author[ISEE,KMI]{A.~Okumura}
\author[Tokushima]{R.~Orito}
\author[INFN]{A.~Rugliancich}
\author[ICRR]{S.~Sakurai\fnref{nowatomu}}
\author[Saitama]{N.~Sasaki}
\author[Saitama]{Y.~Sunada}
\author[Ibaraki]{M.~Suzuki}
\author[Konan]{K.~Tamura}
\author[Yamagata]{J.~Takeda}
\author[Saitama]{Y.~Terada}
\author[ICRR,MPI]{M.~Teshima}
\author[Yamagata]{F.~Tokanai}
\author[Tokai]{Y.~Tomono}
\author[Tokai]{S.~Tsujimoto}
\author[Konan]{Y.~Tsukamoto}
\author[Tokai]{Y.~Umetsu}
\author[Konan]{T.~Yamamoto}
\author[Ibaraki]{T.~Yoshida}

\fntext[nowatdesy]{Now at Deutsches Elektronen-Synchrotron (DESY), 15738 Zeuthen, Germany}
\fntext[nowatpo]{Now at Faculty of Physics, Astronomy and informatics, Nicolaus Copernicus University, Piwnice,  87-148 {\L}ysomice, Poland}
\fntext[nowatchiba]{Now at Chiba University, 1-33, Yayoicho, Inage-ku, Chiba, Japan}
\fntext[nowatritsu]{Now at Research Organization of Science and Technology, Ritsumeikan University, 1-1-1 NojiHigashi, Kusatsu, 525-8577, Shiga, Japan}
\fntext[nowatomu]{Now at Graduate School of Science, Osaka Metropolitan University, Sumiyoshi, Osaka 558-8585, Japan}

\address[ICRR]{Institute for Cosmic Ray Research, University of Tokyo, 5-1-5, Kashiwa-no-ha, Kashiwa, Chiba 277-8582, Japan}
\address[ISEE]{Institute for Space--Earth Environmental Research, Nagoya University,\\Furo-cho, Chikusa-ku, Nagoya 464-8601, Japan}
\address[IFAE]{Institut de Fisica d'Altes Energies (IFAE), The Barcelona Institute of Science and Technology, Campus UAB, 08193 Bellaterra (Barcelona), Spain}
\address[Kyoto]{Department of Physics, Kyoto University, Kyoto 606-8502, Japan}
\address[Konan]{Department of Physics, Konan University, Kobe, Hyogo, 658-8501, Japan}
\address[IAC]{Instituto de Astrof\'isica de Canarias and Departamento de Astrofísica, Universidad de La Laguna, La Laguna, Tenerife, Spain}
\address[Saitama]{Grad. Sch. of Sci. and Eng., Saitama University, 255 Simo-Ohkubo, Sakura-ku, Saitama 338-8570, Japan}
\address[Ibaraki]{Faculty of Science, Ibaraki University, Mito, Ibaraki, 310-8512, Japan}
\address[MPI]{Max-Planck-Institut f\"ur Physik, Boltzmannstr. 8, 85748 Garching, Germany}
\address[KMI]{Kobayashi--Maskawa Institute for the Origin of Particles and the Universe, Nagoya University\\Furo-cho, Chikusa-ku, Nagoya 464-8602, Japan}
\address[Tokushima]{Graduate School of Technology, Industrial and Social Sciences, Tokushima University, Tokushima 770-8506, Japan}
\address[Yamagata]{Department of Physics, Yamagata University, Yamagata, Yamagata 990-8560, Japan.}
\address[INFN]{SFTA Department, Physics Section, University of Siena and INFN, Siena, Italy}
\address[Tokai]{Department of Physics, Tokai University, 4-1-1, Kita-Kaname, Hiratsuka, Kanagawa 259-1292, Japan}
\address[cnrs]{Sorbonne Université, CNRS/IN2P3, Laboratoire de Physique Nucl\'eaire et de Hautes Energies, LPNHE, 4 place Jussieu, 75005 Paris, France}





\begin{abstract}
The camera of the Large-Sized Telescopes (LSTs) of the Cherenkov Telescope Array Observatory (CTAO) consists of 1855 pixels that are grouped into 265 high-performance photomultiplier tube (PMT) modules. Each module comprises a seven-light-guide plate, seven PMT units, a slow control board, and a readout board with a trigger board. 
The requirements for the PMT modules include various aspects, such as photon detection efficiency, dynamic range, buffer depth, and test pulse functionality.

We have developed a high-performance PMT module that fulfills all these requirements. Mass-production and quality control (QC) of modules for all four LSTs of the northern CTAO have been completed.  Here we report on the technical details of each element of the module and its performance, together with the methods and results of QC measurements. 
\end{abstract}


\end{frontmatter}

\section{Introduction}
Imaging Atmospheric Cherenkov Telescopes (IACTs) have been the leading instruments in gamma-ray astronomy above 100 GeV since the discovery of the first TeV gamma-ray source, the Crab Nebula, by the Whipple telescope in 1989 \cite{1989ApJ}. 
Continuous efforts to improve the instruments have led to the detection of more than 200 sources by different generations of telescopes \cite{TeVCat}. The success of IACTs stems mainly from 
using the atmosphere as part of the detector and 
imaging the air showers. When a high-energy gamma ray enters the atmosphere, an electromagnetic cascade called an air shower is initiated. Charged particles in the air shower travel faster than the speed of light in the atmosphere, emitting Cherenkov photons. These photons are spread over an area of $\sim 10^5$ m$^2$ on the ground. Detecting them and imaging the evolution of the air shower is the essential technique of IACTs, achieving a huge effective area with good angular and energy estimation of incoming gamma rays.

The Cherenkov Telescope Array Observatory (CTAO) is the latest-generation gamma-ray observatory, which is expected to have more than one order of magnitude higher sensitivity than any existing IACT arrays. Its energy coverage is from 20 GeV to 300 TeV. Such a wide energy coverage is realized by using three different sizes of telescopes: Large-Sized (LSTs), Medium-Sized (MSTs), and Small-Sized Telescopes (SSTs). Two arrays will be built in both hemispheres. For the northern site, four LSTs and nine MSTs are planned to be built over a 0.25 km$^{2}$ area as a first phase with a possible future extension. For the southern site, 14 MSTs and 37 SSTs will first be built in a few square kilometer area, which may be followed by two or more LSTs and a few tens of additional MSTs and SSTs
\footnote{https://www.ctao.org/emission-to-discovery/array-sites/}.
The LSTs have the largest mirror dish of 23~m in diameter and a parabolic shape. 
These telescopes ensure the sensitivity of the CTAO in the lowest energy range, down to 20 GeV.
In order to achieve high sensitivity at such low energies, in addition to the large mirror area, high-performance cameras with high photon detection efficiency, short pulse width, fast waveform sampling and low noise are required.


We built such cameras using numerous custom-made seven-pixel modules, which include the readout electronics.
In the next section, we will describe these seven-pixel photomultiplier tube (PMT) modules of the LST camera and their performance requirements.

There are slight differences between the PMT modules for the first LST and those for the second to fourth LSTs as described in Section~\ref{sec:LightGuides}, \ref{ssec:PMT-structure}, and \ref{ssec:pmt-performance}.
 Throughout the paper, we refer to the first telescope as LST-1 and the others as  LST-2--4.

\section{PMT module of the CTAO-LST camera}
\label{sect:Module}
\begin{figure}[!htb]
    \centering
    \resizebox{0.5\textwidth}{!}{
      \includegraphics[angle=0,width=5cm]{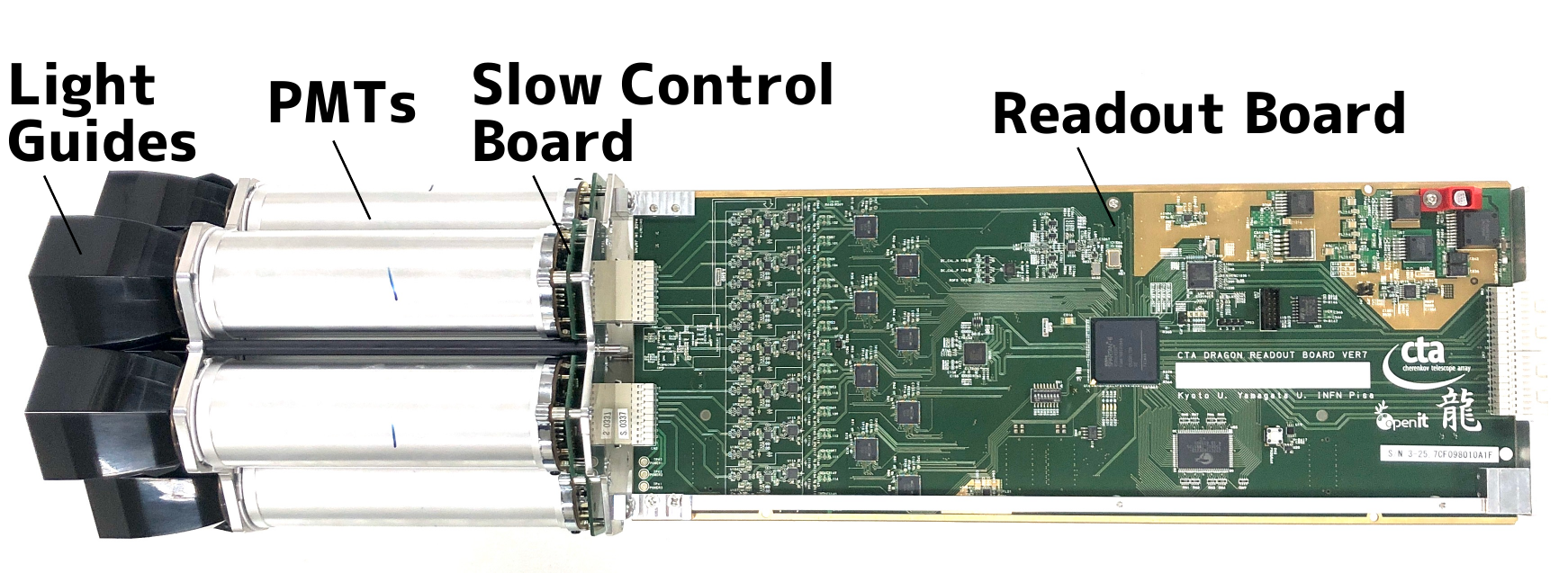}
    }
    \resizebox{0.5\textwidth}{!}{
    \includegraphics[angle=0]{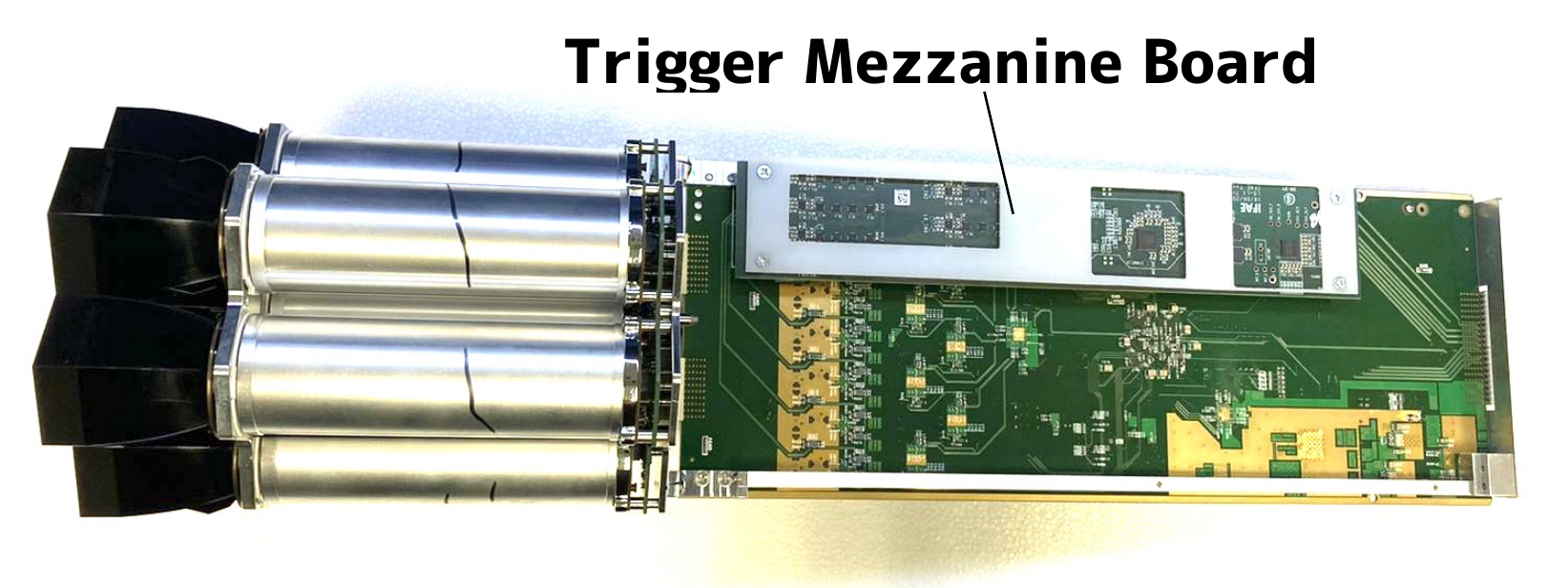}
    }
    \caption{Photos of the PMT module. The front side (top) and the back side (bottom) are shown. 
    The black plastic cones on the left are the LGs seen from the side.
    Metal tubes next to LGs encompass PMT units. Seven PMT units are connected to the SCB, and then the readout board is connected to it. On the back side of the readout board (bottom photo), the trigger mezzanine board, protected with a polytetrafluoroethylene plate, is attached. }    
    \label{fig:module_photo}
\end{figure}

The camera of LSTs of CTAO has 1855 pixels and the size of the light-sensitive area is about 2.2 m in diameter. 
To facilitate maintenance, it is composed of 265 PMT modules, shown in Figure~\ref{fig:module_photo}.
Each PMT module consists of a plate with seven light guides (7 LG), seven PMT units, a slow control board (SCB), a readout board, and a trigger mezzanine board. 
Cherenkov photons focused by the telescope mirror onto the camera are further directed to the photocathode of the PMTs by LGs. The PMTs convert the photons to electrical signals and the readout board records the signal waveforms. The SCB controls and monitors the PMTs, while the trigger mezzanine board generates the triggers based on the signal from the PMTs.

The PMT unit is composed of a PMT, a Cockcroft-Walton (CW) circuit for high voltage (HV) generation, a preamplifier board, and an aluminum tube. The tube shields the other elements. The seven PMT units are connected to the SCB,
which is then coupled to the readout board.
The trigger mezzanine board is attached to the readout board on the backside.

Inside the LST camera, each module is connected to a backplane (BP) board. The adjacent BP boards are connected with each other forming a network, from which the modules receive a 10~MHz clock, a pulse-per-second clock, low-level trigger signals, and readout trigger signal. The trigger signals generated by modules are also sent through the BP network towards the trigger interface board in the camera \cite{TIB}. The digitized waveforms of PMT output are sent to the camera server
through the Ethernet cables connected on each BP board.

 This PMT module has been designed to fulfill the requirements listed below.
\subsection*{Photon detection efficiency}
\begin{figure}[!htb]
\centering
\includegraphics[width=0.45\textwidth]{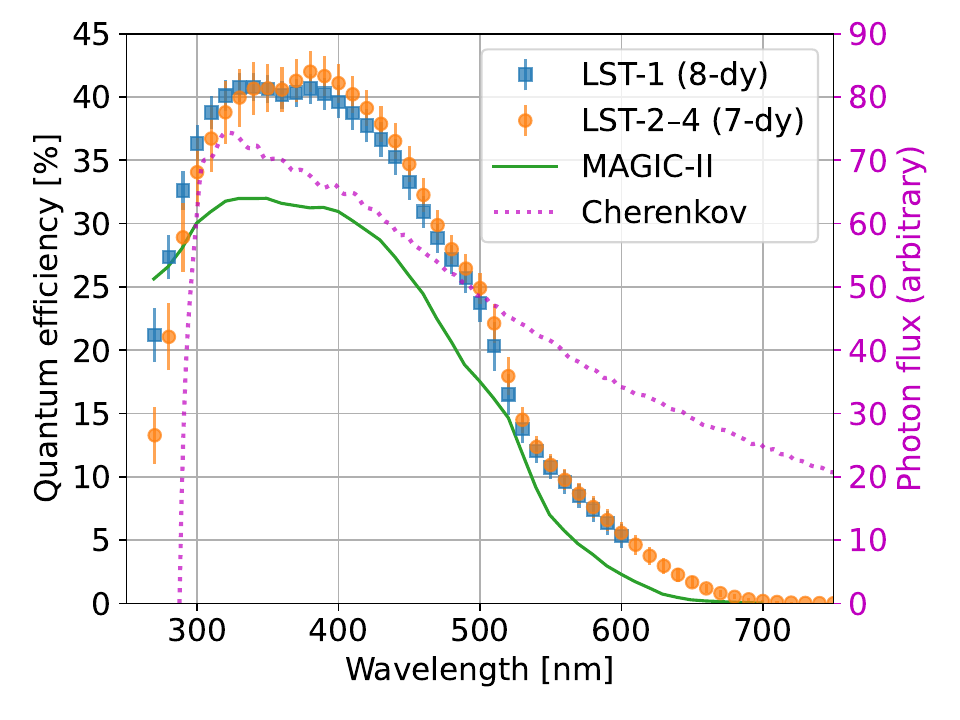}
\caption{Quantum efficiency spectrum of PMT R11920-100 (blue) for LST-1 and R12992-100 (orange) for LST-2--4, respectively, which are averaged over tubes. The former and latter were measured for 270 nm--600 nm and 270 nm--750 nm, respectively. The error bars indicate the standard deviation of the distribution. Solid green curve is that of MAGIC-II PMTs~\cite{TRIDON2010437}. Dotted magenta curve is a Cherenkov photon spectrum simulated for gamma-ray showers.}
\label{fig:PMT-QE}
\end{figure}

The amount of detected Cherenkov photons from air showers 
has a direct impact on telescope performance.
The CTAO requires the average photon detection efficiency of the camera to be greater than
$15$ \% \cite{CTAReq}. Here, the efficiency is weighted by a reference Cherenkov spectrum in the wavelength range 300--550~nm, taking into account the mirror reflectivity, the camera entrance window transmittance, the light guide efficiency, and the quantum and collection efficiencies of photosensors.
The Cherenkov spectrum of simulated gamma-ray showers, with a zenith angle of $20^{\circ}$ and detected at 2200 m above sea level, can be found in Figure~\ref{fig:PMT-QE}, together with the quantum efficiency (QE) of the two types of PMT (see Section~\ref{ssec:PMT-structure}), measured by Hamamatsu Photonics K.K. (HPK)\footnote{https://www.hamamatsu.com/}. Also, the overall optical efficiency of the full Cherenkov telescope to background light, including night sky background (NSB), which is dominated by airglow and zodiacal light~\cite{NSB_LaPalma}, and photons from the ground, must be such that
$\epsilon_{sig} / \sqrt \epsilon_{bg} > 0.35$, where $\epsilon_{sig}$ and $\epsilon_{bg}$ are detection efficiencies of Cherenkov photons and background photons, respectively \cite{CTAReq}. 

\subsection*{Pulse width}

For LSTs, the NSB detection rate is expected to be approximately 250 MHz per pixel\footnote{This corresponds to 250 photoelectrons produced by the NSB photon flux at the PMT photocathode per microsecond per pixel.}. Due to the nature of air showers and the parabolic shape of the LST optics, the arrival times of Cherenkov photons at the camera are spread over only a few to several nanoseconds. To selectively trigger on air showers, the signal pulse width from the photosensor, after pre-amplification, must have a full width at half maximum (FWHM) of less than 3 ns. Furthermore, to minimize the impact of NSB photons on image analysis, this pulse width must be maintained until the signal is recorded. Since variations in PMT manufacturing are inevitable, the requirement is that the average pulse width across all pixels should be less than 3 ns, with no individual pixel exceeding 3.5 ns.

\subsection*{Gain}
CTAO is expected to be operational for more than 20 years.
In order to slow down the degradation of the PMT dynodes (see, e.g.~\cite{HSU2009267,PMThandbook}), the gain of PMTs needs to be kept as low as possible. 
On the other hand, a high signal-to-noise ratio (see {\it Dynamic range and linearity} section below) requires a sufficiently high gain. To balance them, the gain of PMTs is limited to 40000 if the pulse width at this gain is less than 3 ns in FWHM. If not, a larger gain is used and the output signal is electrically attenuated (see Section~\ref{ssec:pmt-performance}). 
It is important to note that we define the PMT gain based on the output charge integrated over a 5-ns time window around the pulse peak, which corresponds to the typical signal integration window used in the air shower analysis.
As the pulse widths slightly differ among PMTs and vary with applied HVs, it is convenient to define the gain for a given time window. The true gain may be slightly higher than the gain defined in this way due to a small fraction of charge (typically less than 10\%) falling outside this window. 

\subsection*{Afterpulsing rate}
Large afterpulses (see e.g. \cite{PMThandbook}) can cause false triggers. A high afterpulsing rate of PMTs may lead to a higher trigger threshold.
Simulations of an LST array indicate that the energy threshold would not worsen if the rate of afterpulsing with an amplitude of $\geq 4$ photoelectrons (p.e.) is lower than $4 \times 10^{-4}$.

\subsection*{Signal sampling rate}
The sampling rate of the signal waveform is higher than 1~GHz, in order to record the shape of the 3 ns-wide pulses. This reduces the impact of NSB photons, which arrive randomly at the camera, on Cherenkov light pulses from the air shower.
\subsection*{Depth of buffer}
\begin{figure}[!htb]
    \centering
    \includegraphics[width=70mm]{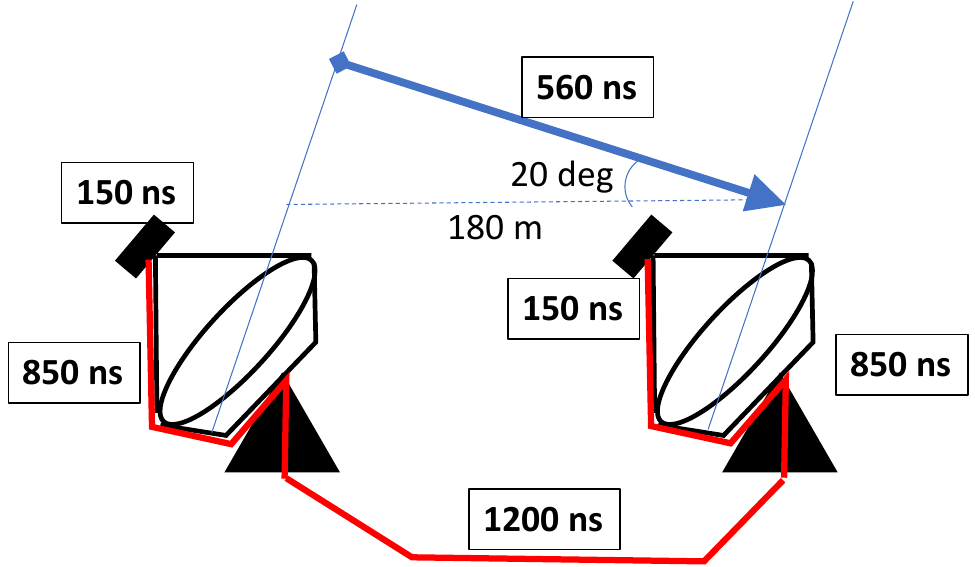}
     \caption{A diagram illustrating the need of 4 \textmu s buffer. Arrival time difference of Cherenkov light between the two telescopes can be as large as 560 ns. It takes roughly 150~ns to transfer the trigger signal inside the camera. Path length of the trigger fibers along the telescope is about 850 ns, while it is up to 1200 ns between the telescopes. }\label{fig:stereo}
\end{figure}

When a trigger signal is issued in one telescope, it is sent via an optical fiber to neighboring telescopes up to $\sim 180$ m apart.
If a trigger issued in the local camera coincides with the one from another telescope, the image is recorded. Taking into account the entire path between two cameras and the decision-making time,
the depth of the buffer holding the waveform must be 4 \textmu s or longer (see Figure \ref{fig:stereo}).

\subsection*{Dynamic range and linearity}

%

To discriminate one-photoelectron signal from electrical noise, 
the readout noise has to be lower than 0.25 p.e., while, to measure the gamma-rays up to 10 TeV, 2000 p.e. per pixel need to be counted.
The linearity should be guaranteed within 10\% over the entire dynamic range.


\subsection*{Configurable trigger setting}

The trigger threshold must be configurable to adapt the data rate to the data acquisition capabilities, depending on sky brightness conditions. For each pixel, the signal amplitude of up to 20 p.e. should not saturate the individual trigger electronics, while the discriminator threshold for the summed signal (see section \ref{sect:mezzanine}) should be configurable up to 100 p.e. level.

\subsection*{Test pulse injection and parameter monitoring}
The PMT module is required to have a test pulse injection function to perform the self-calibration of the readout and the trigger system. Also, it should be able to monitor the temperature, humidity, PMT high voltage, and anode current to ensure  safe and stable operation.
\subsection*{Power Consumption}
The power budget of the camera limits the consumption per pixel to 3 W.

\vspace{1cm}
 In the next section, the LGs are described, while in Section 4, the PMT units are introduced. Section 5 discusses the SCBs, and the readout boards are briefly explained in Section 6. The trigger mezzanine boards are demonstrated in Section 7. Finally, the quality check of the produced PMT modules is discussed in Section 8.

 %

\section{Light Guides}
\label{sec:LightGuides}
\begin{figure}[!htb]
    \centering
    \includegraphics[clip, width=0.45\textwidth]{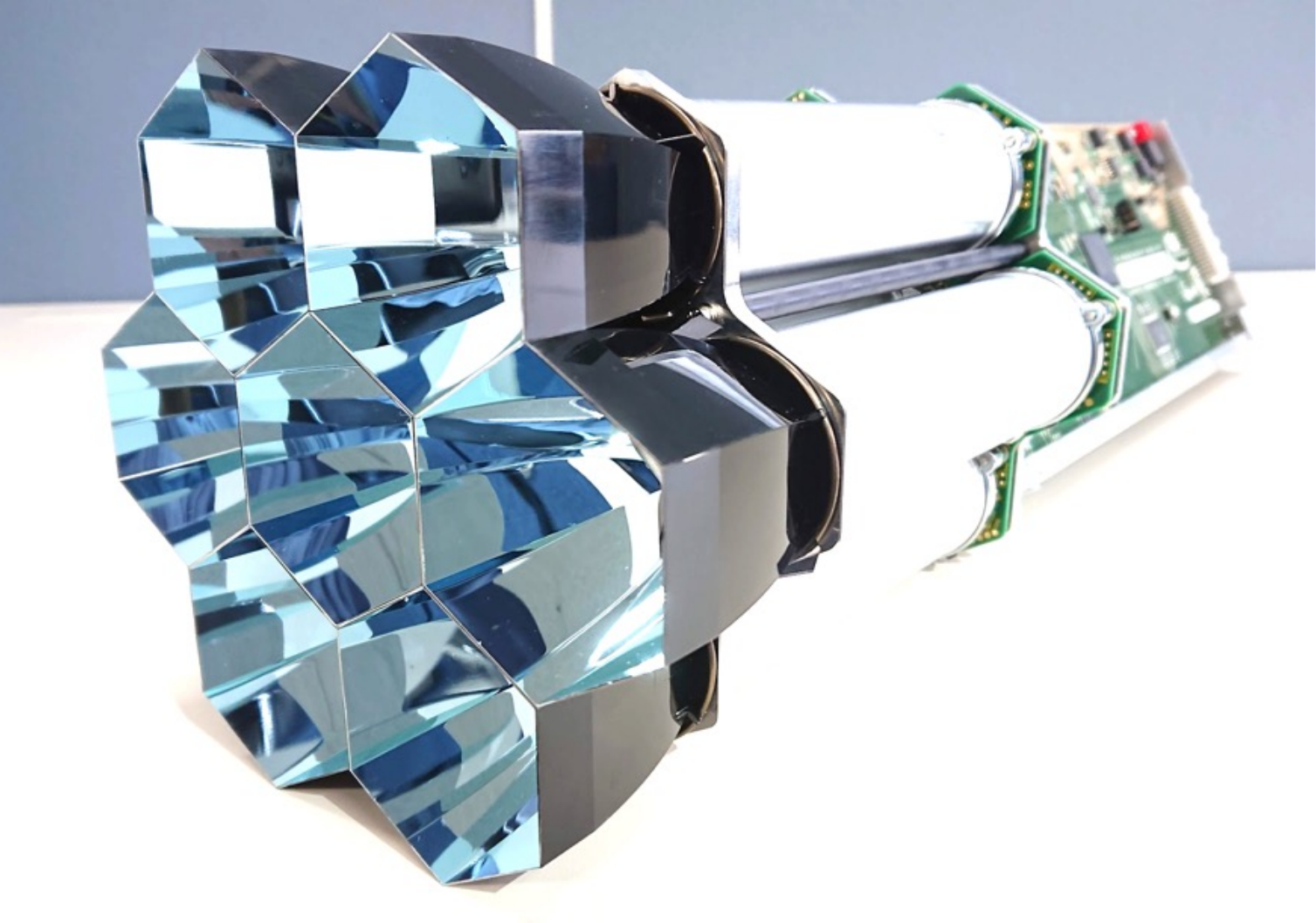}
     \caption{(The photo of the 7-LG plate attached to the PMT module}
     \label{fig:LGphoto}
\end{figure}

\begin{figure}[!htb]
    \centering
      \includegraphics[clip, width=0.35\textwidth]{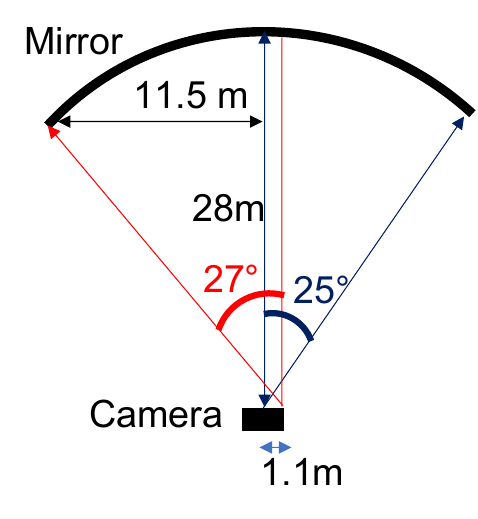}
     \caption{A schematic figure of the required viewing angle of LGs at the center and at the edge of the camera}
     \label{fig:LGView}
\end{figure}

The imaging camera of the LSTs uses 38 mm diameter PMTs arranged in a hexagonal pattern 
with a 50 mm pitch.
Non-imaging light guides (LGs) are necessary to reduce the dead area between these PMTs and to reduce the albedo.
The LGs are designed as hexagonal-shape frustums with a flat-to-flat distance of 50\,mm at the entrance and 25\,mm at the exit (see Figure \ref{fig:LGphoto}).
To accommodate the viewing angle of the main mirrors at the center of the camera, which is $\pm25$ degrees, and the maximum angle at the camera edge, which is $\pm27$ degrees (see Figure \ref{fig:LGView}), the LGs need to have an acceptance angle of $\pm29$ degrees, which includes a 2-degree tolerance.
Impinging Cherenkov photons are first reflected from the segmented mirrors of the LST optics and focused on the focal plane. They are then concentrated onto the PMT photocathodes by the interior mirror surfaces of the LGs.

The shape of the reflective surface of LGs is designed using a B{\'e}zier curve and optimized to enhance photon detection efficiency using the ROBAST ray-tracing simulation library~\cite{Okumura_2017,Okumura2012, OKUMURA201638}.

The LGs are composed of 
acrylonitrile butadiene styrene resin and are manufactured through injection molding by Kyoei Engineering Corporation\footnote{https://kyoeieng.co.jp/}. 
To make the specular surfaces highly reflective and to keep the surface finish uniform, the metal mold we used for plastic injection was processed through ultra-precision machining.
The micro-roughness of both the plastic and coating surfaces was kept below 5 nm.
The reflective coating is comprised of three layers: aluminum, \ce{SiO2}, and \ce{Ta2O5}.
This coating was optimized to increase the reflectance of ultraviolet light with a wavelength of 320 nm for a wide range of incident angles. A low-temperature deposition was performed for this coating by Tokai Optical Corporation\footnote{https://www.tokai.com}. 

The top edge of the frustum is 0.3\,mm thick, which creates approximately 2\% of dead space in the focal plane. Seven LGs are fixed onto a 5\,mm thick aluminum plate by using latching mechanisms. On the other side of this plate, the photocathodes of the seven PMTs are secured using springs that press the PMT onto the aluminum plate. Since the PMT cathode has a voltage of around $-$1000\,V, a 1.5\,mm gap is included between the PMT surface and the reflective coating to prevent conduction and discharge. 

The light collection efficiencies of these LGs have been measured as 
a relative anode sensitivity (RAS, refer to \cite{Okumura_2017}  for more details).
The RAS was measured for different wavelengths ranging from 310 to 650\,nm for different incident angles from 0 to 40 degrees.

\begin{figure}[!htb]
    \centering
    \subfigure[Incident angle dependence of RAS for LST-1 LGs]{\includegraphics[clip, width=0.75\linewidth]{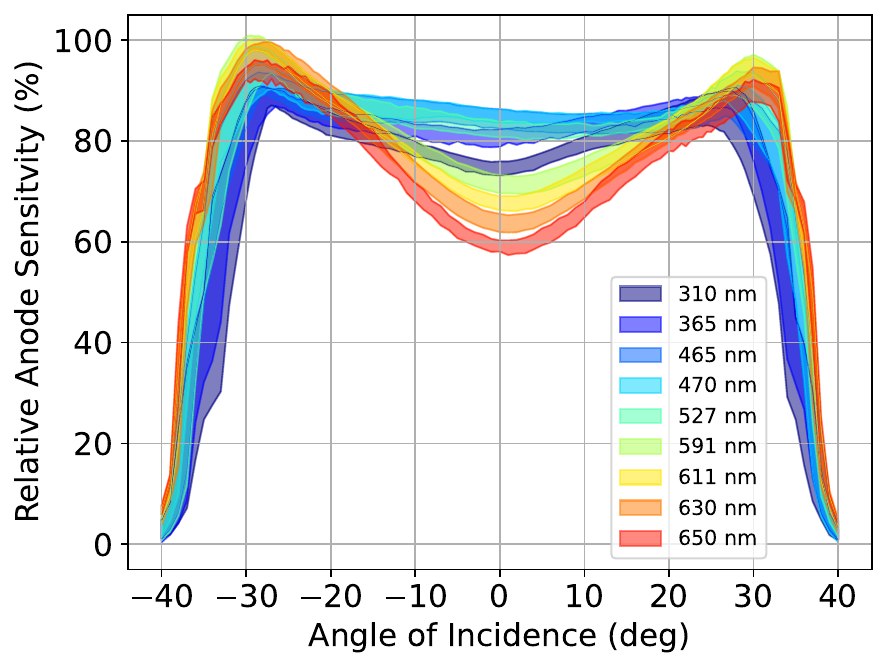}}
     \subfigure[Incident angle dependence of RAS for LST-2-4 LGs]{\includegraphics[clip, width=0.75\linewidth]{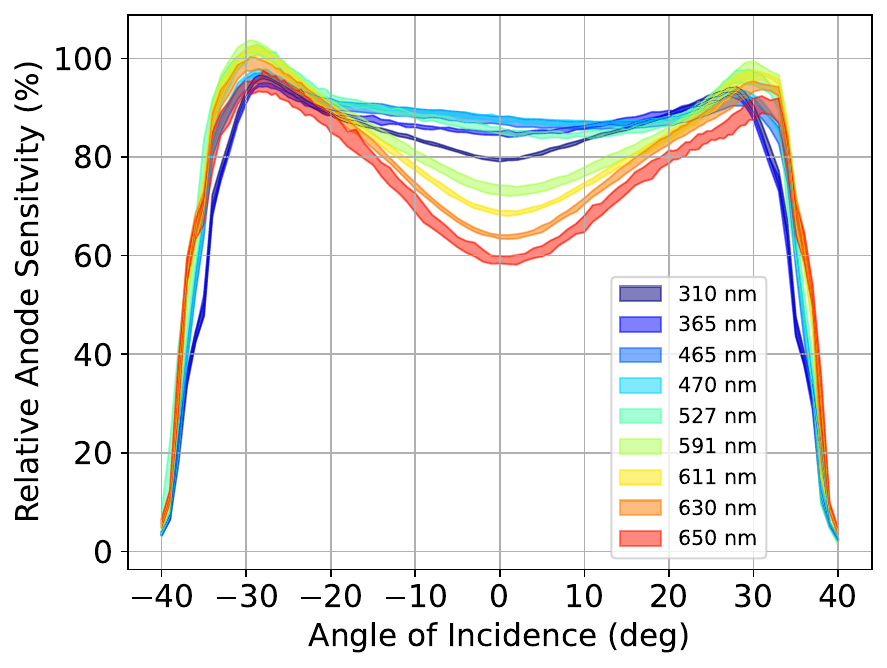}}
     \subfigure[Wavelength dependence of RAS]{\includegraphics[clip, width=0.75\linewidth]{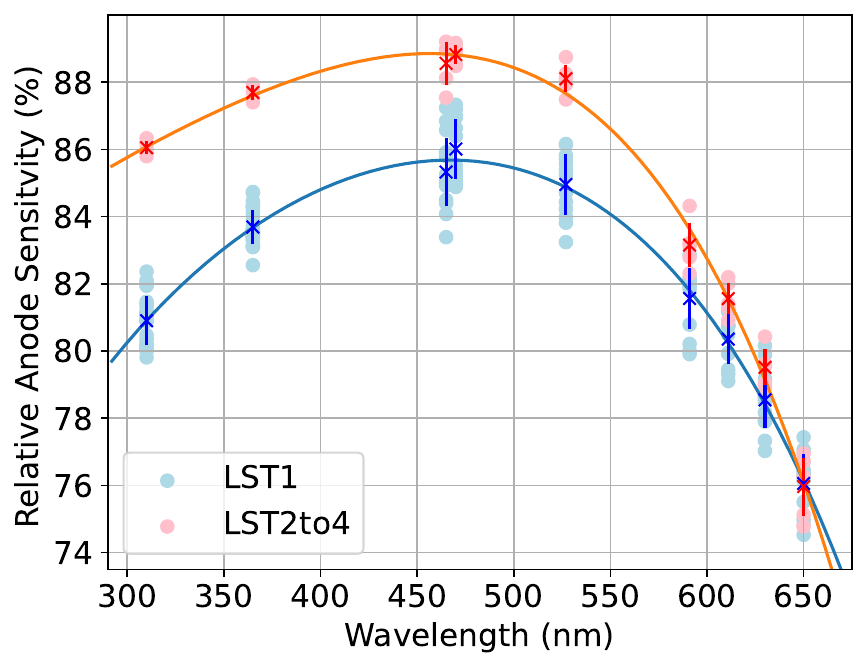}}
    \caption{Relative anode sensitivities (RASs) measured in a laboratory. RAS is regarded as the collection efficiency of the LGs if the angular and position dependence of the PMT-anode sensitivity is negligible \cite{Okumura_2017}. 
    In the panel (a) and (b),  
    RASs of five LGs for LST-1 and four LGs for LST-2--4 are plotted by shading within the range between the measured minimum and maximum values, as a function of the incident angles of the light. 
    Each color of the shading in this figure corresponds to the 9 wavelengths of the incident light from 310~nm to 650~nm, as shown in the legend. 
    The panel (c) shows the expected averaged RAS accounting for the angular distribution of the incident light from the main mirror \cite{Okumura_2017} as a function of wavelength of the incident light. Blue points and curve are for LST-1 and red points and curve are for LST-2--4, as indicated in the legend. The error bars represent the standard deviations between measurements.}
%
%
    \label{fig:Lg}
\end{figure}

As shown in Figure~\ref{fig:Lg}(a), the angular dependence of RAS gradually changes with the wavelength.
This behavior can be attributed to the fact that the reflectance of the dielectric coating on the LGs inherently depends on both angle and wavelength.
Note that the peak structures at $\pm 30$ degrees are due to the angular dependence of the QE of LST PMTs \cite{Okumura_2017} .

Figure ~\ref{fig:Lg}(b) compares the averaged RAS of LST-1 LGs and LST-2–4 LGs as a function of wavelength. For each wavelength, angular dependence was averaged out taking into account the angular distribution of the segmented mirrors.
It exceeds 80\% within the range of 300 to 550\,nm. The light guides for the LST-2--4 have a slightly improved collection efficiency due to an adjusted deposition procedure. The total collection efficiency of the light guide
is estimated considering the distributions of incident angles and the wavelength of Cherenkov light at the altitude of the observation site between 300 and 550 nm. It is 84\% for LST-1 and 88\% for LST-2--4. 

\section{PMT Unit} 
\label{sec:PMT}
\begin{figure}[!htb]
	\centering
        \includegraphics[clip, width=0.45\textwidth]{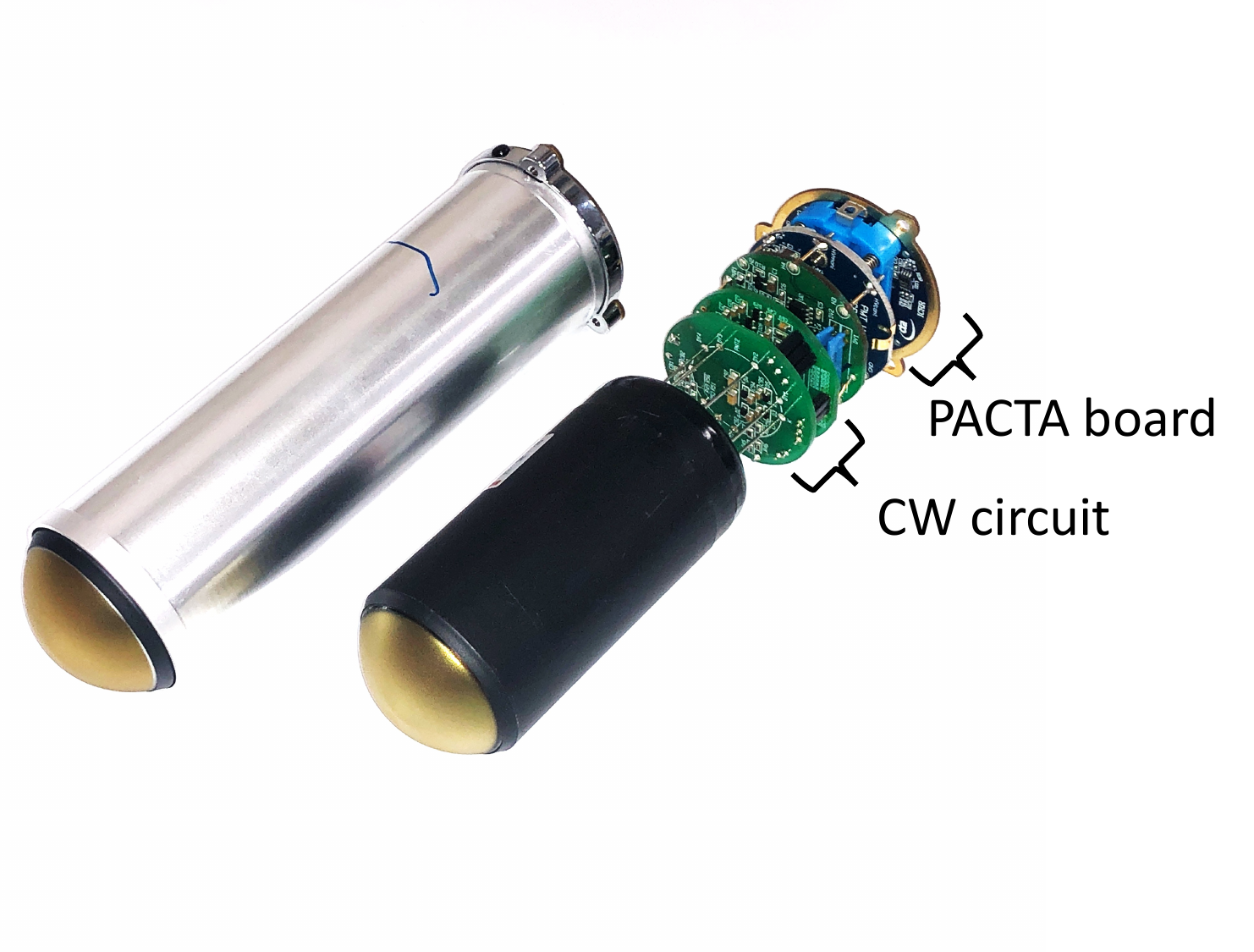}
	\hspace{0.01\textwidth}	
	\caption{Photo of PMTs with and without the aluminum tube. The Cockcroft-Walton circuit for HV production and the pre-amplifier board called PACTA are seen.}
 \label{fig:PMTphoto}
\end{figure}

The PMT unit is composed of a PMT, a PACTA (PreAmplifier for Cherenkov Telescope Array) board, and a CW circuit board~\cite{CW1932,CW-HPK}. These are assembled together and placed inside an aluminium cylinder to protect them from electromagnetic noise (see Figure \ref{fig:PMTphoto}). Aluminium makes the cylinder light and easy to mass-produce.
\subsection{Components}
The details of each component are provided below.
\paragraph{PMT}
\label{ssec:PMT-structure}
The LST camera is designed to detect low-energy events that emit only faint Cherenkov light. To achieve this, the PMT needs to have a high QE and a reduced afterpulsing rate. The LST-1 camera uses the R11920-100 PMT, which was developed by HPK together with the authors of~\cite{MIRZOYAN2016640,MIRZOYAN2017603}.
It features a hemispherical super-bialkali photocathode~\cite{NAKAMURA2010276} and a matte entrance window, which enhances the QE~\cite{MIRZOYAN2006230,TOYAMA2015280}.
HPK also adopted a dynode structure from another PMT, R8619, which has a very low afterpulsing rate~\cite{2013ICRC...33.1178T}. The glass bulb is surrounded by a conductive paint (HA treatment) and mu-metal to reduce the effect of the external electric and magnetic field, respectively~\cite{PMThandbook}. 

For LST-2--4, HPK developed the R12992-100 PMT with further performance improvements.
R11920-100 and R12992-100 have the same mechanical dimensions. The diameter of the tube is 1.56 inch (=39.6 mm). 
While R11920-100 has eight dynodes (8-Dy), R12992-100 has seven (7-Dy). The number of stages was reduced to increase the lifetime (limited by the degradation of dynodes --- see Section~\ref{sect:Module}), maintaining a gain of 40000. More specific information can be found in Section~\ref{ssec:pmt-performance}.

\paragraph{CW circuit}
The CW circuit boards were developed by HPK. They supply a negative HV of ${\sim} 1000$--$1500$ V between the photocathode and the anode. The voltage can be controlled via the SCB (see Section~\ref{ssec:SCB-PMT}), specifically to adjust the PMT gain. However, the voltage between the photocathode and the first dynodes is fixed to 350~V by a Zener diode. This configuration stabilizes the number of electrons produced at the first dynode, independently of variations in the total applied voltage. Consequently, this ensures that the event-by-event relative fluctuation of the signal response is almost independent of the gain. Therefore, once this fluctuation factor is established, one can monitor the PMT gain using a stable light source~\cite{1997ICRC....7..265M} .

\paragraph{PACTA board}
\label{sec:pacta}
The PACTA board converts the PMT output current to a voltage signal using a fast transimpedance amplifier application specific integrated circuit (ASIC) called a PACTA~\cite{PACTA}.
The board has an AC coupling with a capacitor to filter slow components of the PMT signal (the slow current is transfered to SCB and digitized there. See Section \ref{sect:SCB}).
The PACTA amplifies the signal by a high gain with a transimpedance of $1200\,\Omega$ and a low gain with that of  $80\,\Omega$. They are output through different channels. Both of them have a bandwidth of about $\SI{400}{\mega\hertz}$ at $\SI{-3}{\decibel}$, and the power consumption is less than $\SI{150}{\milli\watt}$ when operating at $\SI{3.3}{\volt}$.
Right before the PACTA chip input, there is a controllable switch that allows us to select either the PMT signal or a test pulse generated in the SCB as an input signal. In addition, a digital temperature sensor is mounted, which is read by the SCB as described in Section~\ref{ssec:SCBmonitor}. 
 
\subsection{Performance}
\label{ssec:pmt-performance}
In this section, we describe the typical performance of the PMT units in terms of some important characteristics. The results of the characterization performed for each of them are reported in Section~\ref{sec:QC}.
\paragraph{Pulse shape}
It is well known that in a PMT, the higher the applied voltage, the shorter the output pulse width \cite{PMThandbook}.
The FWHM of the PMTs is required to be $< \SI{3}{ns}$ during operation in order to prevent the NSB photons (In LSTs, the photoelectron rate is ${\sim} 250$ MHz per pixel) from causing a trigger due to coincidental pile-up. However, when we apply a sufficiently high HV to set the width $< \SI{3}{ns}$, the gain exceeds 40000 for many PMTs. For such cases, we attenuate the PMT signal by implementing resistors on each PACTA board to dissipate a portion of the output current as heat. 
The higher gain compensates for this attenuation resulting in an output charge equivalent to the gain of 40000 while maintaining a pulse width less than 3 ns in FWHM. 
The attenuation factors are 3.5 or 2 depending on each specific R11920-100 tube.
For R12992-100, we asked HPK to reduce the number of dynodes from eight to seven, thereby making the gain at the same HV smaller than R11920-100. The attenuation factor was then reduced to 3 or 1.7. This change is expected to lengthen their lifetime.

\begin{figure*}[!htb]
	\centering
	\includegraphics[width=\textwidth]{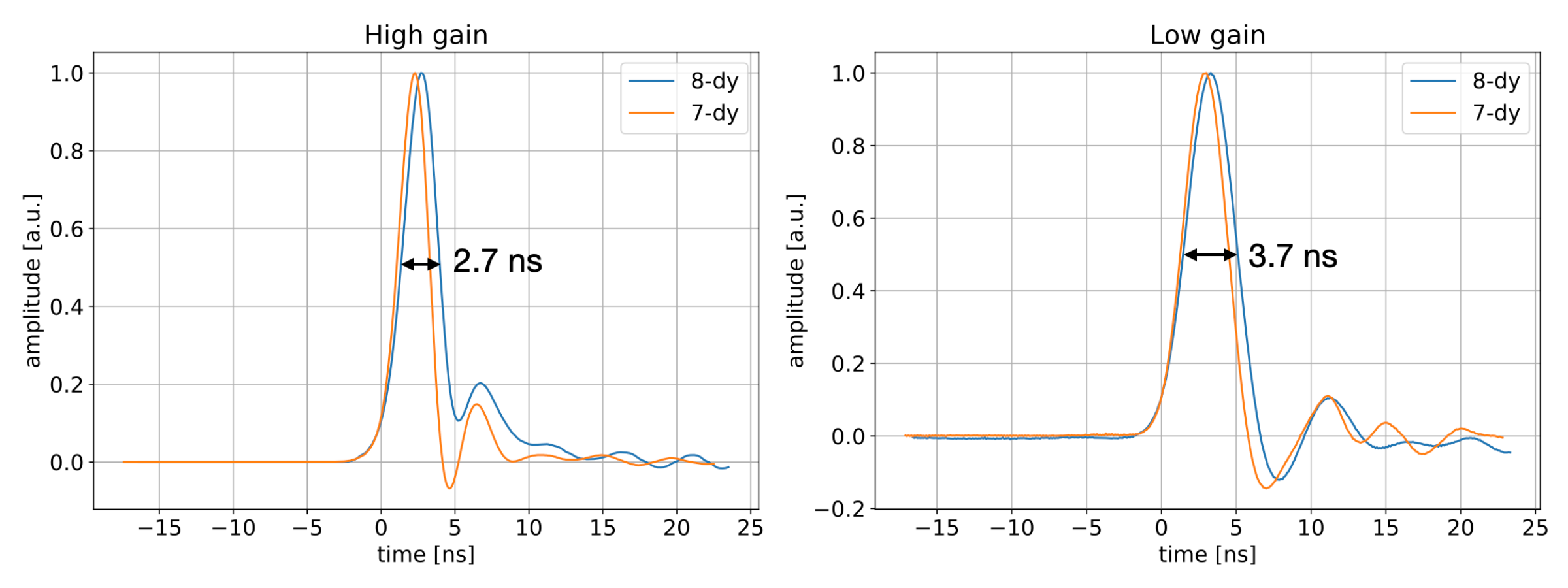}
	\caption{Typical pulse shape of PMT R11920-100 (Blue; 8-Dy, ${\sim} 1060$~V applied) for LST-1 and R12992-100 (Orange; 7-Dy, ${\sim} 1140$~V applied) for LST-2--4; the left and right plots shows an output waveform through the high gain channel and the low gain channel of PACTA, respectively. The ringing after the main peaks is due to the AC coupling of the signal output.}
	\label{fig:PMT-PulseShape}
\end{figure*}

In Figure~\ref{fig:PMT-PulseShape}, typical pulse shapes of R11920-100 and R12992-100 with a gain $\approx 40000$ are shown (${\sim} 1060$~V for R11920-100 and ${\sim} 1140$~V for R12992-100). These signals go through the PACTA and are measured at the readout board. For this measurement, the sampling frequency of the readout board was 5 GHz. They were measured with a pin-hole mask in front of the PMT to reduce the transit time spread due to the difference in the arrival point of photons on the photocathode. Owing to the signal attenuation leading to a higher voltage for the gain of 40000, 
The pulse width (FWHM) through the high gain channel of PACTA is ${\sim} 2.7$ ns for R11920-100 and ${\sim} 2.3$ ns for R12992-100. It should be noted that the pulse widths in low gain channels are larger, specifically ${\sim} 3.7$ ns for R11920-100 and ${\sim} 3.3$ ns for R12992-100. This discrepancy is primarily due to the lower bandwidth of the low gain in the readout board, a consequence of the more extended electrical wiring. However, this poses no issue for the analysis of detected showers since the effect of NSB on large signals ($>60$ ~p.e.), for which low gain channels are used, is negligible.

\begin{figure}[!htb]
	\centering
        \includegraphics[clip, width=0.45\textwidth]{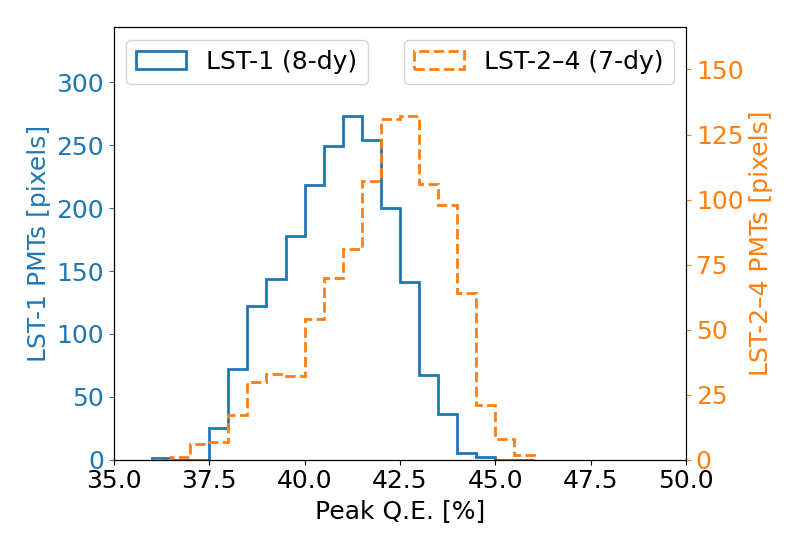}
	\hspace{0.01\textwidth}	
	\caption{Distribution of the peak QE of PMT R11920-100 (blue; 1987 tubes) for LST-1 and R12992-100 (orange; sampled 1000 tubes) for LST-2--4}
 \label{fig:PMT-PeakQE}
\end{figure}

\paragraph{Quantum efficiency}
The QE values that we use for the QC and Monte Carlo simulations of the instrumental response were supplied by HPK for a subset of tubes that are installed on the telescopes. The spectrum and the peak distribution are depicted in Figures~\ref{fig:PMT-QE} and~\ref{fig:PMT-PeakQE}, respectively.  
On average, the peak QE is 40.8\% for R11920-100 and 42.0\% for R12992-100. These values are much higher than ${\sim} 32$\% seen for a reference model tube from MAGIC-II~\cite{MIRZOYAN2007449,TRIDON2010437}. 
R11920-100 and R12992-100 were produced in different assembly lines leading to a minor difference in their QE.
Combined with the mirror reflectivity (approximately 90\%), the LG collection efficiencies and the loss at the camera entrance window (approximately 10\%), the Cherenkov photon detection efficiencies of the LST cameras between 300 and 550~nm are 26\% for LST-1 and 27\% for LST-2--4, largely exceeding the CTAO requirement (see Section~\ref{sect:Module}). Also the Cherenkov photon detection efficiency between 250 and 700 nm is $\epsilon_{sig}$ = 0.21 for LST-1 and 0.22 for LST-2--4, while that for NSB in the same range is $\epsilon_{bg} $= 0.11 for both, fulfilling the requirement $\epsilon_{sig} / \sqrt{\epsilon_{bg}} > 0.35$ (see Section~\ref{sect:Module}).

\paragraph{Afterpulsing}
In the specification, the rate of afterpulsing with an amplitude of $\geq 4$ p.e.\,shall be lower than $2\times 10^{-4}$. 
The delivered PMTs satisfied the above condition; the measurements provided by HPK as well as by~\cite{TOYAMA2015280,MIRZOYAN2016640,MIRZOYAN2017603,2021NIMPA100765413T} confirmed it to be $(0.1$--$1.2)\times 10^{-4}$. This rate is more than one order of magnitude lower than those of the reference MAGIC-II PMTs~\cite{HSU2009267}, $(2$--$8) \times 10^{-3}$. 

\begin{figure}[!htb]
    \centering
    \includegraphics[clip, width=0.45\textwidth]{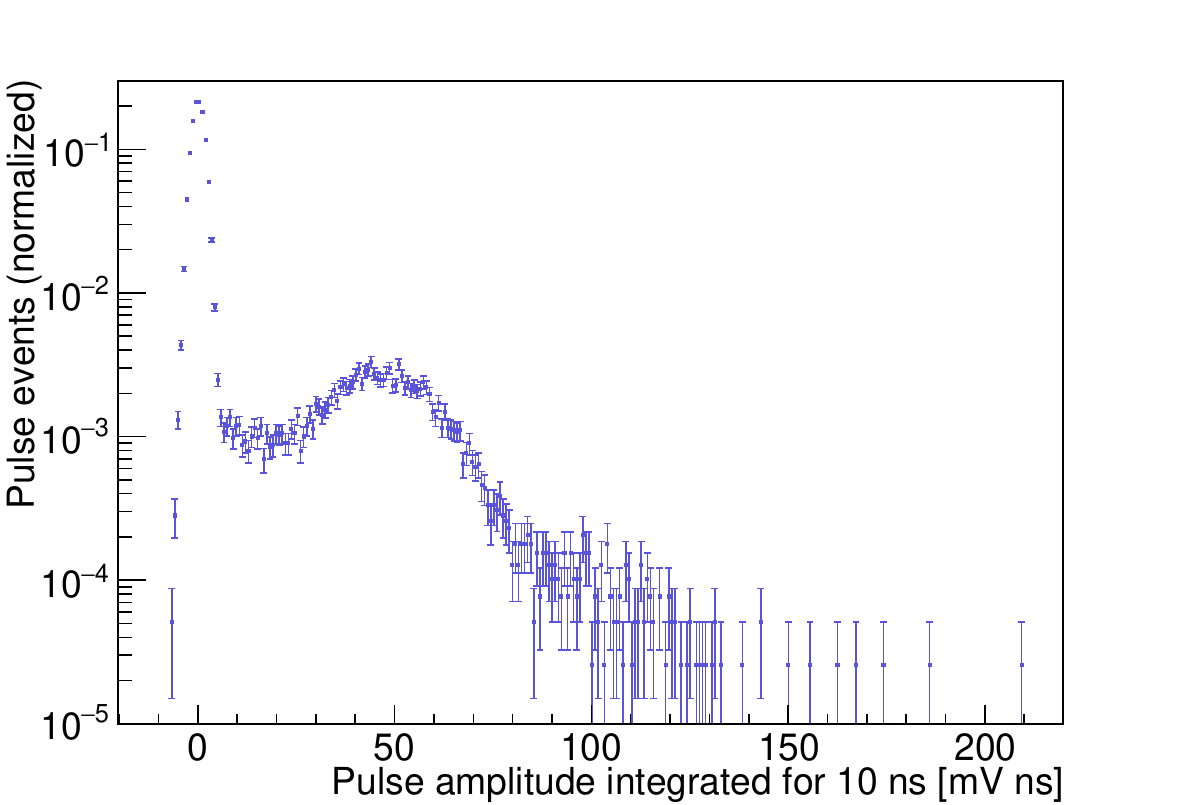}
    \caption{Measured distribution of the time-integrated amplitude of 55002 events taken from one of the R11920-100 PMTs for LST-1. The peaks of 0 p.e.\,, 1 p.e.\,and 2 p.e.\,can be seen around $\SI{0}{mV\cdot ns}$, $\SI{50}{mV\cdot ns}$, and $\SI{100}{mV\cdot ns}$, respectively. The error bars show the $1\sigma$ statistical uncertainty.}
    \label{fig:observed_sphe-spectrum}
\end{figure}

\paragraph{Single photoelectron response}
The response to a single-p.e.\,input is one of the important characteristics for the photon intensity resolution and the efficient trigger. 
 One needs to measure the single-p.e.\,response spectrum and the excess noise factor (ENF) precisely to implement them in the analysis and simulations. The ENF (denoted by $F$ in this paper) represents how much the photodetector output fluctuates with respect to that of the input photoelectrons.
In the laboratory, we measured the single-p.e.\,response for several PMTs by a method similar to that illustrated in~\cite{TAKAHASHI20181}.
First, we applied an HV of 1500~V to a PMT to obtain a high signal-to-noise ratio. Then, we illuminated it with more than 50000 faint light flashes, with the average light content of ${\sim} 0.2$ p.e. (It should be noted that precise light intensity was not required). The signal waveform was integrated over a fixed time window of 10 ns, which was sufficient to include most of each pulse accounting for the jitter in the pulse timing. On average, 1 p.e.\,was detected out of 6 flashes. The output charge histogram $N(q_i)$, where $q_i$ is the charge corresponding to the $i$-th bin, displayed a few peaks. The first peak is the 0 p.e. charges, and the second one is the single p.e. charges. The 2-p.e.\,peak is also visible, but the 3- and more p.e.\,components are negligible as demonstrated in Figure~\ref{fig:observed_sphe-spectrum}.

To estimate the single p.e.\,response solely as a function of the charge $q$, we modeled the 0- and 2-p.e.\,components and subtracted them from $N(q_i)$ as follows. 
First, we modeled the 0-p.e.\,peak by a Gaussian function $m_0(q)$, which should represent the fluctuation of the readout noise. It is normalized so its integral equals unity. We also introduced an empirical model $m_1(q)$ for the true normalized distribution of the single p.e.\,output charge without readout noise. To represent the main peak and the small-charge tail, the model consists of the King function $m_1^{King}$ \cite{King_1983,Ackermann_2013} and an adaptive power-law function $m_1^{APL}$, which are here defined as 
\begin{equation}
\label{eq:King}
m_1^{King}(q) = \frac{a}{\sigma^2}\left(1-\frac{1}{\gamma} \right) \left[ 1+\frac{1}{2\gamma}\left( \frac{q-\mu}{\sigma} \right)^2 \right]^{-\gamma},
\end{equation}
and
\begin{equation}
\label{eq:AdaptPL}
m_1^{APL}(q) = 
  \begin{cases}
    b \left(1-q/s\right)^{\alpha+(\beta-\alpha)q/s} & \text{if $q\le s$,} \\
    0                 & \text{if $q>s$,} 
  \end{cases}
\end{equation}
where $a, \mu, \sigma, \gamma, b, s, \alpha$, and $\beta$ are fit parameters. 
Then, a convolution of $m_0(q)$ and $m_1(q)=m_1^{King}(q)+m_1^{APL}(q)$ gives us a noise-convolved 1-p.e.\,peak model:
\begin{eqnarray}
M_1 (q) = m_0(q) \circledast m_1(q).
\end{eqnarray}
Similarly, the output charge distribution of 2 p.e.\,should be
\begin{eqnarray}
M_2 (q) = m_0(q) \circledast m_1(q)\circledast m_1(q)
\end{eqnarray}

In addition, the number of events with 0, 1, and 2 p.e.\,input should follow a Poisson distribution with the mean value $\lambda$. 
Given the low intensity of the light flashes, we considered the contribution from 3 p.e.\,events to be negligible. We then fitted the measured charge distribution $N(q_i)$ with
\begin{eqnarray}
M(q) = N_0 \left( P^\lambda_0  m_0(q) + P^\lambda_1  M_1 (q) + P^\lambda_2  M_2 (q) \right),
\end{eqnarray}
where $N_0$ is the total number of events and $P^\lambda_k$ is the Poisson probability to find $k$ when the mean is $\lambda$. 
The value of $\lambda$ was restricted to be within $\pm 2$\% of the ratio $\langle N\rangle/\langle M_1\rangle$, which should ideally equal the mean of the incident photoelectron count per pulse. 
Finally, we obtained the most realistic single-p.e.\,response $\hat{N}_1(q_i)$ by 
\begin{eqnarray}
\hat{N}_1(q_i)=N(q_i) - N_0 \left( P^\lambda_0  m_0(q_i) + P^\lambda_2  M_2 (q_i) \right).
\end{eqnarray}
This procedure accounted for possible residuals in the data after the fit rather than relying solely on the model $m_1(q)$.

For a complete QC of the PMT modules reported in Section~\ref{sssec:OperationVoltage}, we employed a simpler method with multi-Gaussian fitting to estimate the mean of the single-p.e.\,response.

\begin{figure}[!htb]
    \centering
    \includegraphics[clip, width=0.45\textwidth]{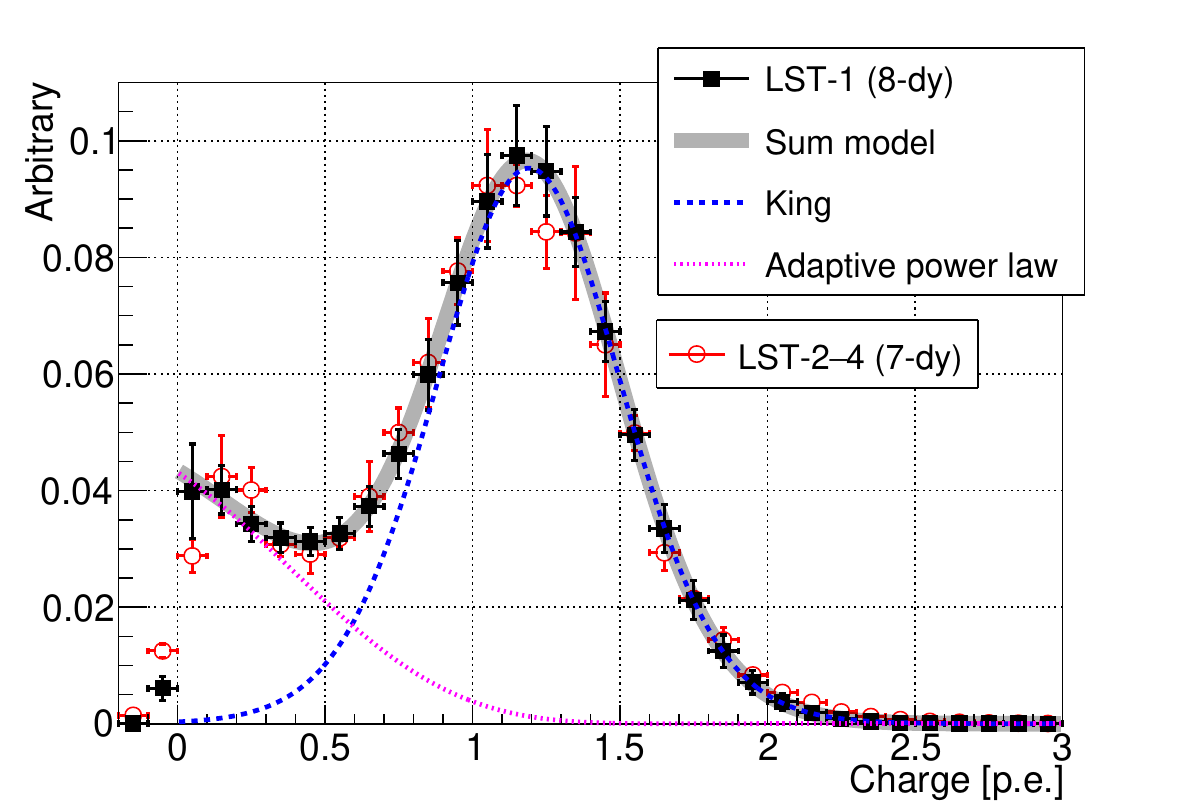}
    \caption{Black squares (Red circles): Average of the single-photoelectron (p.e.) response of 307 R11920-100 PMTs for LST-1 (five R12992-100 PMTs for LST-2--4) as a function of the output charge. They were estimated on the basis of our measurement. Their charge values are scaled so that the mean of the distribution from each PMT equals unity. Error bars: Standard deviation of the distribution of those tubes. Gray bold curve: Modeled single-p.e. response spectrum of PMT R11920-100. Blue dashed curve: Component of the King function (Equation~\ref{eq:King}) of the model. Magenta dotted curve: Component of the adaptive power-law function (Equation~\ref{eq:AdaptPL}) of the model.}
    \label{fig:modeled_sphe-spectrum}
\end{figure}

Figure~\ref{fig:modeled_sphe-spectrum} shows the average and standard deviation of $\hat{N}_1(q_i)$ of 307 R11920-100 tubes and five R12992-100 tubes, derived by this method. We also plotted $M_1(q)$, fitted with the average data points of the R11920-100 tubes in the charge range from 0 to 2.5 p.e.
The measured points for the R12992-100 tubes show a ``cliff'' around 0.1 p.e.\,, but we do not consider this feature to be real as the subtraction of the pedestal component outnumbering the residual 1-p.e.\,one introduces relatively large systematic uncertainty. This is especially significant for the R12992-100 tubes, which have a lower gain than the R11920-100 tubes with the same high voltage. 

We estimated the ENF of each measured tube using $\hat{N}_1(q_i)$. It is formulated by
\begin{equation}
F= \sqrt{1+ \frac{var\left(\hat{N}_1(q_i)\right)}{\langle \hat{N}_1(q_i)\rangle^2}},
\end{equation}
where $var\left(\hat{N}_1(q_i)\right)$ is the variance of $\hat{N}_1(q_i)$~\cite{1997ICRC....7..265M}. Once the value of $F$ is obtained, it is usable to calibrate the gain of the photodetector during observation.
The mean and the standard deviation of $F$ are $1.107\pm 0.009$ for R11920-100 and $1.104 \pm 0.006$ for R12992-100. These numbers are consistent with those measured by \cite{MIRZOYAN2017603,2021NIMPA100765413T}.
This induces an inevitable fluctuation, for example, of ${\sim} 1.5$\% in the photoelectron number estimation for $1000$ p.e., but it is smaller than the Poisson fluctuation.
We estimate the systematic uncertainty due to the region of $<0.2$ p.e.\,to be $\sim 1$\% in $F$ by taking the difference between the fit model and the averaged measured distribution.

\section{Slow Control Board}
\label{sect:SCB}
The SCB interfaces the seven PMT units and the readout board. The main functionalities of the SCB are the following:
\begin{itemize}
    \item Transferring a fast PMT signal to the readout board
    \item Providing power and controlling HVs for PMTs
    \item Monitoring the HVs, low voltages (LV), anode currents, temperature and humidity
    \item Generation of electrical test pulses
\end{itemize}
\subsection{Fast signal transfer}
The output signal from the PMT unit has a FWHM of 2.0 to 3.5 ns. The SCB needs to transmit these fast pulses to the readout board without degradation. This is done by using the printed circuit board of the SCB, which is designed as a six-layer interstitial-via-hole (IVH) board. The signal lines are surrounded by the ground layers on the top and the bottom, together with many IVHs on both sides of the lines in order to reduce the interference from other elements and ensure the signal integrity along the transmission line.

\subsection{Providing power and controlling the HV for PMT units}
\label{ssec:SCB-PMT}
\begin{figure}[!htb]
    \centering
    \includegraphics[width=85mm]{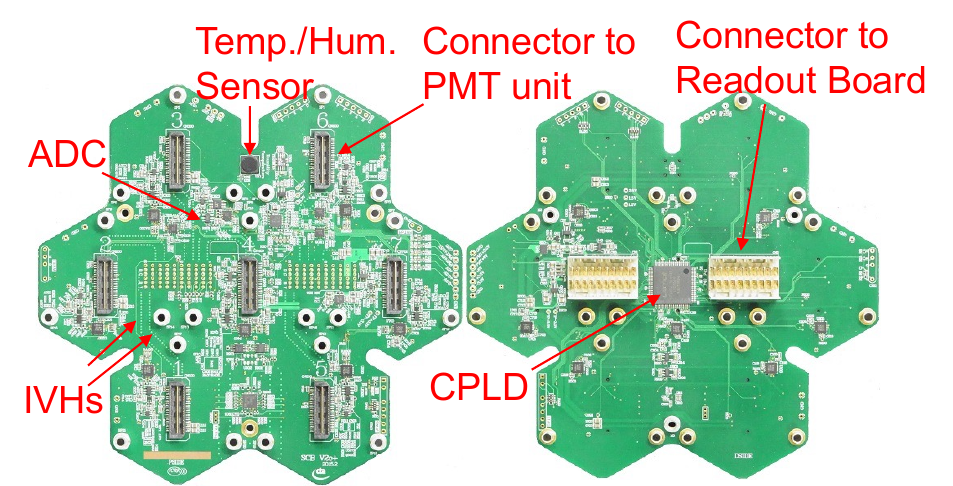}
     \caption{Photos of the SCB. Both the PMT side and the readout board side are shown.}\label{fig:SCB}
\end{figure}

As described in Section \ref{ssec:PMT-structure}, the HV for the PMT is produced with the CW circuit, which requires a 6~V supply voltage and a control voltage. The supply voltage is produced in the readout board, and the SCB distributes it to the seven PMT units. The control voltage for CW (0--1.5 V) is produced on the SCB using digital-to-analog converters (DACs) controlled through a Complex Programmable Logic Device (CPLD, Lattice:LCMXO2-1200ZE-1TG100I) chip on the SCB (see Figure \ref{fig:SCB}). The CPLD communicates with the FPGA on the readout board described in Section \ref{sect:readout} via the Serial Peripheral Interface communication.

\subsection{Monitoring the HVs, LVs, currents, temperature, and humidity.}
\label{ssec:SCBmonitor}
On each PACTA board, a digital temperature sensor is implemented. The value can be read out through 1-wire communication from the CPLD of the SCB. The SCB has also a temperature and humidity sensor, whose values are read out through I$^2$C communication from the CPLD. The applied HVs and anode currents can also be monitored. One-thousandth of the applied HVs are input to the analog multiplexer on the SCB, and the seven values are digitized one by one by the ADC in the SCB. The values are read out via  I$^2$C communication. The anode currents of the seven PMT units are converted to voltages in the PACTA board and then monitored in the same way as HVs, together with several LVs of the board (+3.3 V, $-3.3$ V, +6 V, +1.2\ V ).

\subsection{Test Pulse Injection}
\label{sec:TPI}
\begin{figure}[!htb]
    \centering
    \includegraphics[width=60mm]{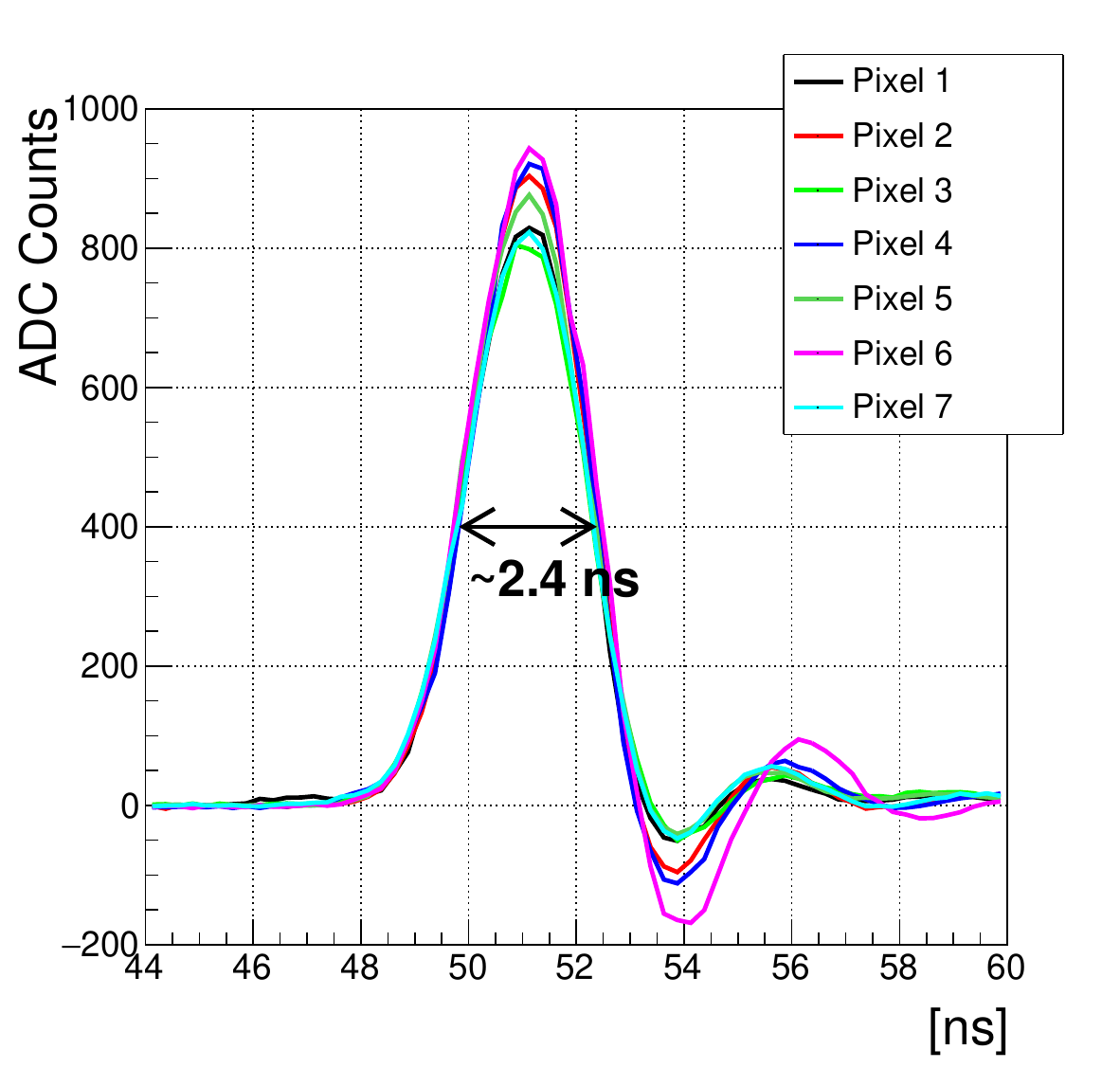}
\includegraphics[width=60mm]{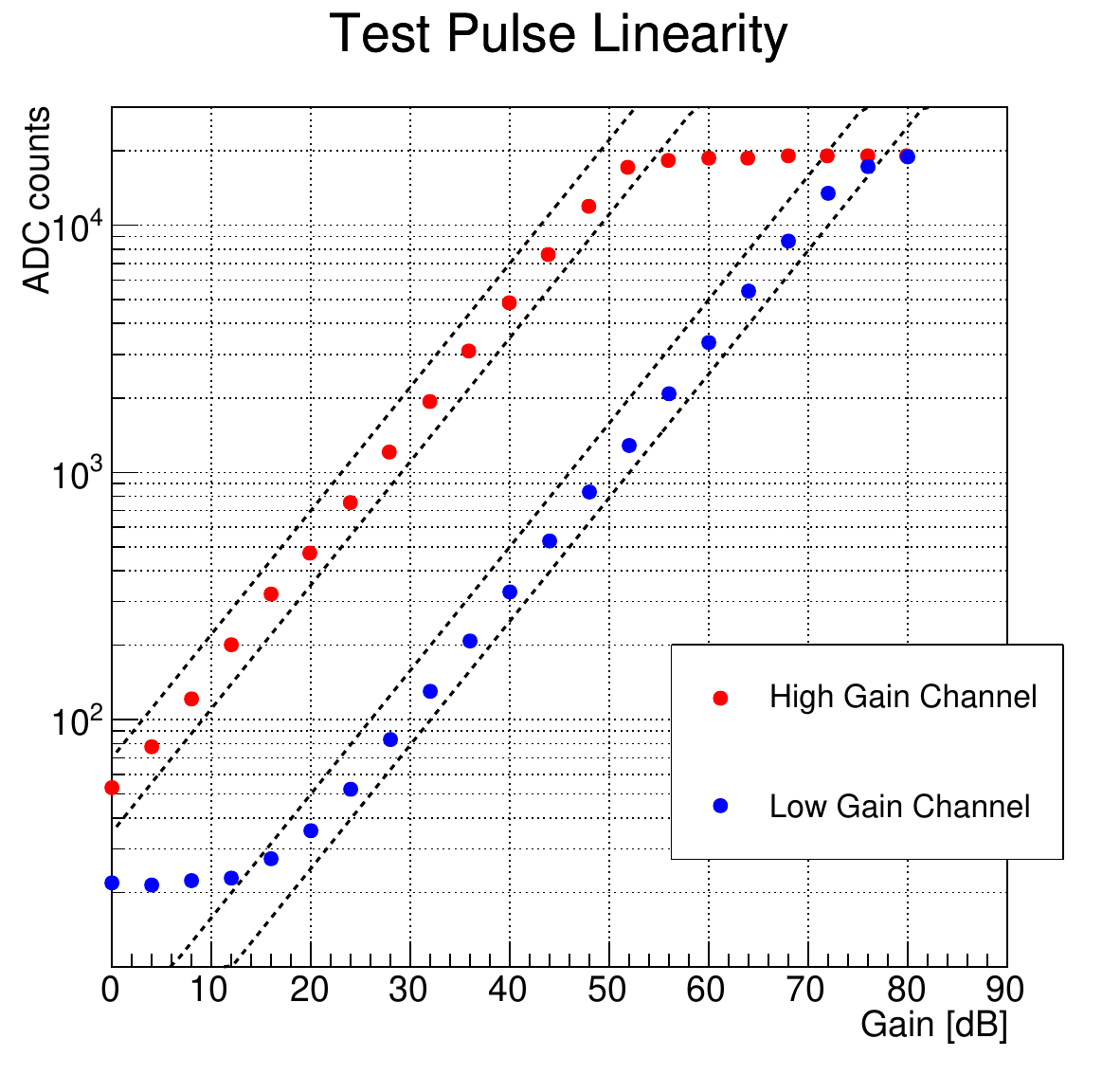}
  \caption{{\it Top}:Pulse shape of test pulses injected from the SCB to the readout board. FWHM is about 2.4 ns.{\it Bottom} Linearity of the test pulse. It can be injected to both high and low gain readout channel. The amplitude can be varied by 4 orders of magnitude (80 in dB). }\label{fig:TestPulse}
\end{figure}

The SCB can generate and inject electrical test pulses in each pixel line, which can be used for linearity checks of the readout board, trigger calibration, data acquisition test, and more. The width of the pulse is around 2.4 ns, which is similar to the PMT signal, as shown in Figure \ref{fig:TestPulse}. Such a fast pulse is generated using a fast logic gate chip (SY55851A) and duplicated rectangular pulses, with one of the pulses delayed. The amplitude can be adjusted over a range of 4 orders of magnitude, from 0.4 to 4000 p.e. with a step size of 1 dB.



\section{Readout board}
\label{sect:readout}

The differential outputs from seven PMT units with two different gains are transmitted to the readout board. On the board, high gain signals are further divided and routed to the readout path and the trigger processing path (see Section~\ref{sect:mezzanine}). 
Those signals are amplified with main amplifiers (ADA4927, \cite{amp}) by factors of 4.0, 5.2, and 19 for low gain, high gain, and trigger path, respectively. The amplification factors are chosen such that the readout noise of high gain channel corresponds to 0.25 p.e. and the saturation amplitude of the low gain channel correspond to 3000 p.e.. With these factors, the high gain channel saturates at 100 p.e., while readout noise of low gain channel is about 8 p.e. ensuring the overlap between high and low gain more than one order of magnitude. It is worth noting that there is a factor 24 difference in amplification at the PACTA board between the high and low gain channels. 
For the trigger branch, the amplification factor was decided to increase the single p.e. amplitude well above the noise in the level-0 (see section \ref{sect:mezzanine}) circuit, 
while avoiding saturation up to 20 p.e..

 The waveform of PMT signals of LSTs needs to be sampled at 1 GHz or higher to suppress the contamination of NSB light in the Cherenkov signal images. In addition, as it was mentioned in Section \ref{sect:Module}, a buffer depth of $4$~\textmu s is also required to record images based on the coincidence trigger between two or more telescopes, which are 100 to 200 meter apart (see Figure \ref{fig:stereo}).

To achieve such a high sampling rate with a large buffer without consuming large power,
the readout board for the LST adopted Domino Ring Sampler 4 (DRS4) chips~\cite{DRS4}, which are based on the use of switched capacitor arrays.
The DRS4 chip was originally developed for the MEG experiment (an experimental project that seeks to observe a muon decay into an electron and a gamma ray\footnote{https://meg.icepp.s.u-tokyo.ac.jp/}) and has been used for other current Cherenkov telescopes~\cite{MAGIC_DRS, FACT_DRS}.
Each DRS4 chip has eight input channels, each consisting of a ring buffer of 1024 capacitors. 
For the LST application, the sampling rate is set to 1.024 GHz, and hence, one ring buffer covers to 1.0 \textmu s. 
To achieve the needed 4 \textmu s buffer, four input channels of the DRS4 chip are used with a cascade mode \cite{DRS4} for a single PMT readout, where the four capacitor arrays sample the input waveforms sequentially.
Therefore, one DRS4 chip can handle only two PMTs.
Eight DRS4 chips are mounted on the readout board. Four of them are used for high gain readout of 7 PMTs and clock signal sampling, and other four are for low gain readout and clock signal sampling.

DRS4 chips continuously sample the signal provided by the PMTs.
When a readout trigger arrives on the board (see Section \ref{sect:mezzanine}), sampling is stopped and the sampled waveform is digitized at 33.3~MHz with an external eight channel 12-bit successive approximation register ADC (AD9637, \cite{ADC}). The dynamic range of the readout is from 0.25 to 3000 p.e. The readout window is set to 40 samples, corresponding to 39 ns. 
The bandwidth of the high gain channel was measured to be 300 MHz.
The digitized signals are buffered within a field programmable gate array (FPGA, Spartan-6 \cite{FPGA}) and formatted together with additional counter information for each event. Those data are transferred to the data acquisition server through a BP board and network switches. Such a data transfer is enabled by an implemented Silicon Transmission Control Protocol ~\cite{Uchida2008} (SiTCP) technology on FPGA. This also enables slow control via User Datagram Protocol. Thus, each subsystem attached to the readout board can be configured through an FPGA on the readout board. The total power consumption including the SCB and 7 PMTs is about 20.2~W during data taking, fulfilling the requirement of 3~W per pixel.

 More detail of this readout board will be published in \cite{DragonPaper}.



\section{Trigger Mezzanine board} 
\label{sect:mezzanine}

As described in Section \ref{sect:readout}, 
the DRS4 chips in the readout board continuously sample the PMT signal. The camera trigger system indicates when the sampled signal should be digitized. The LST uses a three stage analog trigger system developed for CTAO \cite{AnalogueTrigger}.
The first two stages (level-0 and level-1 triggers), which form the trigger decision system for the LST camera, are implemented in the trigger mezzanine board. The level-0 trigger adds the analog signals coming from the seven pixels of each module. The level-1 trigger adds the level-0 triggers for a patch of up to three neighbour PMT modules and checks if the sum is above a given threshold. The third stage is
not part of the PMT module and it is used to implement 
a coincidence trigger between multiple LSTs.

The level-0 trigger is implemented using both analog delay lines and a level-0 trigger ASIC~\cite{L0ASIC}. The former are needed to compensate the dispersion of the PMT transit times, which leads to the signal from the seven pixels not arriving at the same time to the trigger system. 
Monte Carlo simulations show performance degradation 
when misalignments exceeds 0.5~ns. The analog delay lines cover up to 6~ns delays with steps of 0.25~ns. It is worth mentioning that in terms of trigger performance the time alignment is important only among the neighbouring pixels. A time difference of several ns between distant pixels is not relevant if the time alignment is fulfilled among neighbours. Once the signals are time-aligned, they are fed to the level-0 trigger ASIC~\cite{L0ASIC}, where those signals are conditioned and then analogically added together. Copies of the output signal of the level-0 trigger ASIC are distributed, through the BP network, to the neighbor modules.
They are then fed into the level-1 trigger, which is based on a single ASIC~\cite{L1ASIC}.
Inside the ASIC, input level-0 signals are summed up and then examined whether the amplitude of the summed signal is higher than the discriminator threshold. The analog bandwidth at both L0 and L1 stages are around 500 MHz.

The trigger decision system is highly configurable with several parameters that can be set through the slow control and it extends over the full field of view of the LST camera. The flexibility of the trigger provides the capability to reach optimal performance for different observation conditions as for instance different NSB levels. 
It also helps to achieve a homogeneous response across the entire trigger range, meaning that every image should have the same chance to produce trigger regardless of where exactly it hits the camera.

The level-0 and level-1 trigger ASICs are subjected to a specific QC before they get mounted in the mezzanine boards. The mezzanine boards with those ASICs also undergo additional QC before they are plugged into the readout board. The QCs aim to check that the system performs as required for all possible settings as well as the system homogeneity across the full camera. An additional validation is done after the mezzanine is plugged into the readout board to ensure the trigger system is operative in the complete PMT module.

For the mezzanine board QC done in the laboratory an Agilent 81110A function generator was used. It produces a pulse with a FWHM as the one delivered by the PMT through the high gain path (Figure~\ref{fig:PMT-PulseShape}). Pulses of different amplitudes covering the full voltage range are provided independently to both the level-0 and the level-1 triggers. The trigger mezzanine board is configured through the slow control to verify that delay lines and both ASICs operate within the required tolerance. 
All qualified mezzanine boards with calibrated discriminator thresholds have the same efficiency for signals that would correspond to 10 to 50 p.e. 
This ensures the homogeneity of the trigger decision system across the camera. 

\begin{figure*} [!htb]
    \centering
      \includegraphics[clip,width=0.99\linewidth]{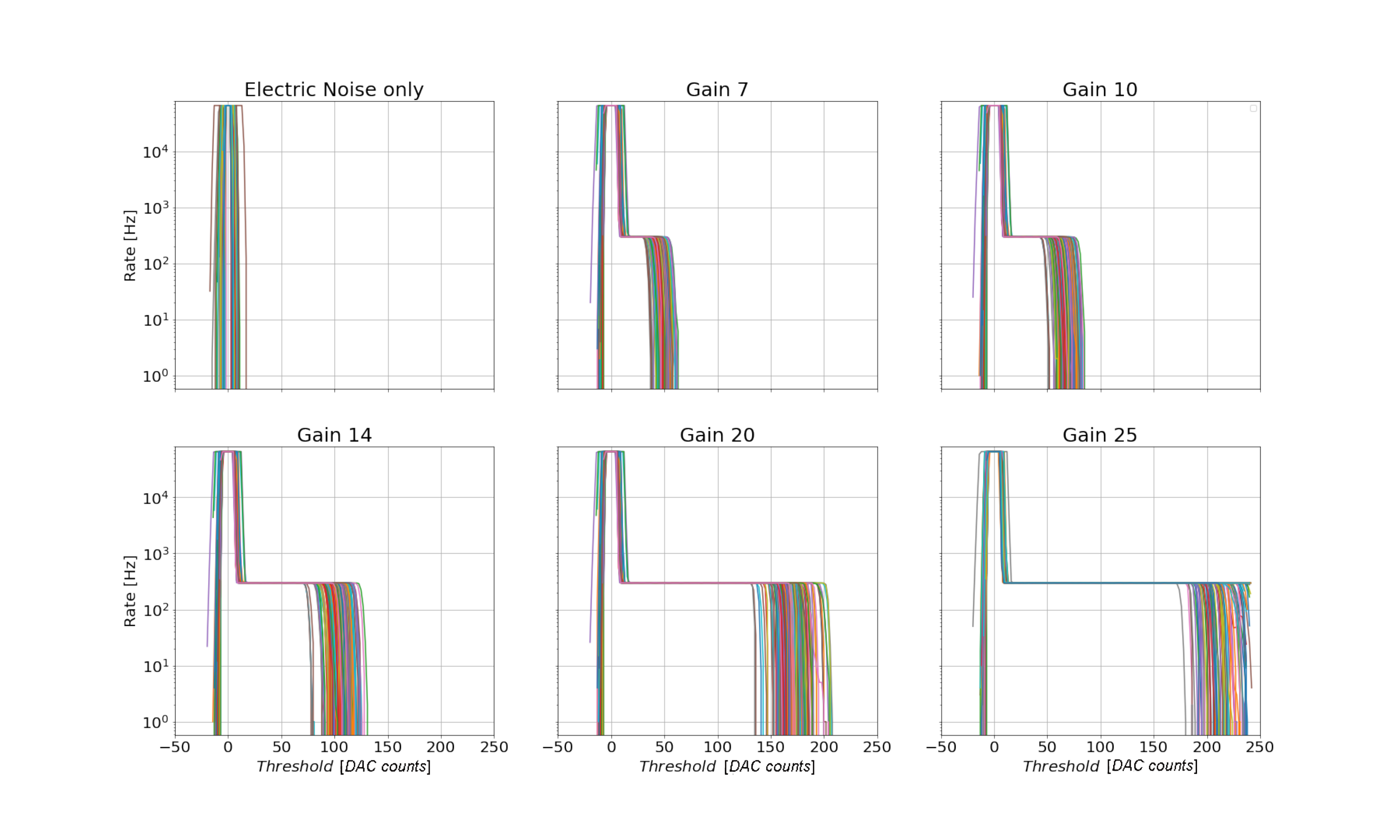}
    \caption{Trigger rate as a function of the level-1 discriminator (L1 rate scans) for increasing amplitudes indicated by increasing gain of the test pulse (see section \ref{sec:TPI}) as well as with only electronic noise. In each panel, the L1 rate scans from all mezzanine boards using only the level-0 signal from its module are shown with different color lines. The horizontal axis is normalized such that the only electric noise measurement peaks at 0 for all mezzanine boards.}
    \label{fig:DTL1_A_local_3.6}
\end{figure*}

Once the modules are installed in the camera, the test pulse injection (see Section~\ref{sec:TPI}) is used to verify that the discriminator of each level-1 can be set 
to provide a homogeneous trigger across the camera. In particular, the trigger rate for different amplitudes of the test pulse is recorded while changing the level-1 discriminator threshold (Figure~\ref{fig:DTL1_A_local_3.6}). 


    \label{fig:DTL0_elN&g20_3.6}


\section{Module Quality Control}
\label{sec:QC}
We conducted a QC of the PMT modules to validate their integrated functions and performance after assembly. Each module forms a highly independent compact unit that converts light into an analog electric signal and subsequently digitizes it. Hence, they can be tested very efficiently once a light source, a BP board, and a power supply are set up. We constructed such a system that acts as a mini-camera, capable of measuring 19 modules simultaneously. Completing one measurement cycle of the 19 modules took roughly 4 hours ($\sim 13$ minutes per module).
The results reported below are based on these mini-camera tests except for the afterpulsing of the LST-1 modules, which was obtained through an alternative setup. This particular setup is described in \ref{sec:PMTQC}.

\subsection{Setup}
\label{sect:qcsetup}

\begin{figure*}[!htb]
    \centering
    \includegraphics[width=0.9\textwidth]{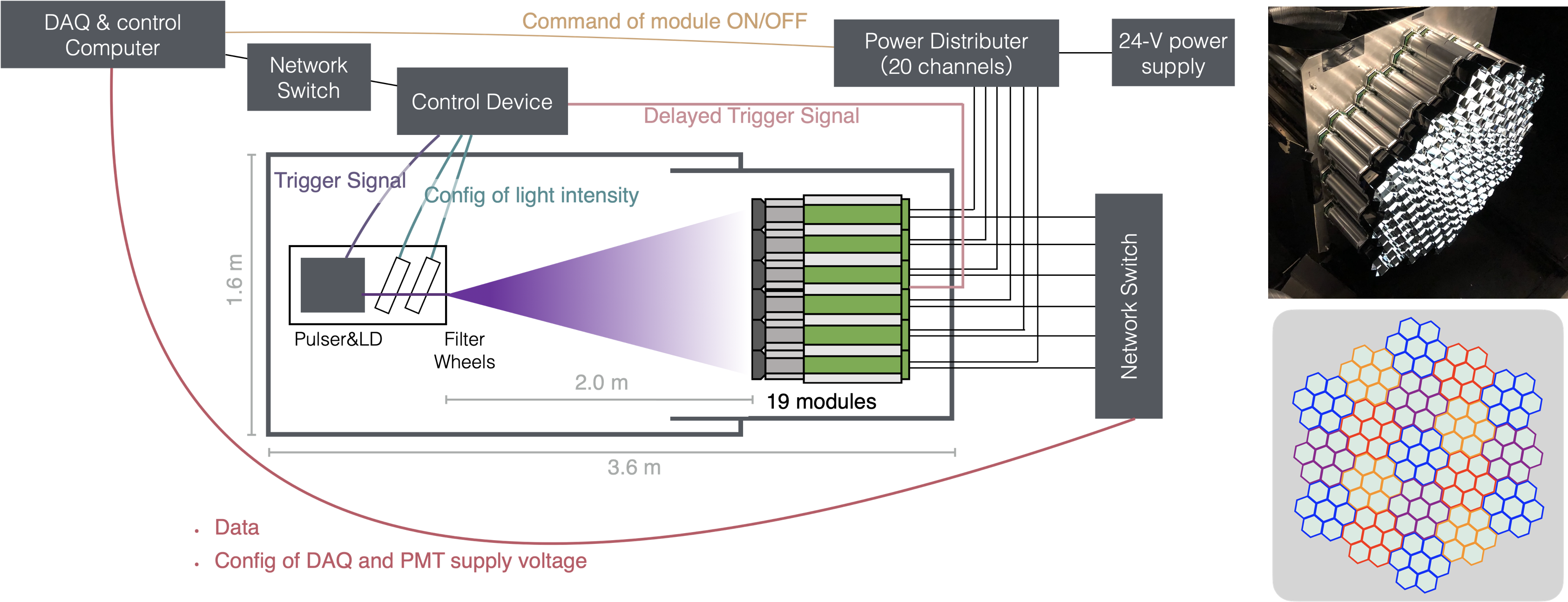}
    \label{fig:QC-MiniCam-front}
    \caption{{\it left}: Overview of the components of the Mini-camera setup and their communication. 
    {\it top right}: Picture of the module rack with 19 modules mounted.
    {\it bottom right}: Physical arrangement of 19 modules held by the rack in a front view. The pixels that are assembled into one module are edged with the same colour.}
    \label{fig:QC-setup}
\end{figure*}

The mini-camera setup is illustrated in Figure~\ref{fig:QC-setup}. The 19 modules were held by a module rack and connected to 19 BP boards.
We used a 405 nm laser diode (LD), Nichia NDV4212, as a light source. The LD was driven by a custom driver circuit to generate very fast light pulses with FWHMs of 800--920 ps~\cite{8533096}.
It was positioned approximately two meters away from the PMTs. 
To achieve a roughly uniform illumination across all pixels, we employed a polytetrafluoroethylene plate to diffuse the LD light. The light flux variation across the pixels then turned out to be a factor of two based on the PMT outputs. For LST-1 measurements, the average laser intensity per pixel was at the $\sim 1000$ p.e.\,level. For LST-2--4, we improved the driver electronics, and the intensity increased by a factor of $\sim 3$.
To configure the light source intensity across \text{more than} three orders of magnitude, we used two neutral-density optical filter wheels. Each of them was equipped with five filters of varying opacities along with an empty slot, allowing for a total of 36 filter combinations. Before starting the QC measurements, the relative opacities of these filters were calibrated by averaging the PMT charge output from 19 modules with the LD light pulsing. For the higher-opacity combinations, we used the output of the high-gain channel because the low-gain output suffers from electrical noise. Conversely, for the lower-opacity ones, we utilized the low-gain to avoid the high-gain output saturating for large photon pulses.

Each PMT output was read out by the assembled readout board. Data acquisition was initiated by a trigger signal that was synchronized with the LD pulser. This signal was delayed by about one rotation of the ring sampler of the DRS4 chip, around 4~\textmu s, to include the light pulse in the time region to be digitized. The 10-MHz clock signal (see Section~\ref{sect:Module}) was also provided by an external unit. The delay and clock signals were injected into the BP network and then distributed to all modules. This process is similar to what happens in the real cameras.

The computer, which was the same model as the camera servers for the LST observations, was used for data acquisition. The data were sent from every module to the computer via a network switch. 
To extract the response to each light pulse from the stored data, we integrated the waveform within a 5-ns time window centered at the peak, consistent with our gain definition (see Section~\ref{sect:Module}).
The power supply to each component, the light intensity, the data sampling configurations and the trigger configurations were controlled from the same computer.



\subsection{Measurements and results}
\subsubsection{Gain-vs.-voltage slope}
\label{sssec:gain-analysis}

\begin{figure}[!htb]
    \centering
    \includegraphics[width=0.45\textwidth]{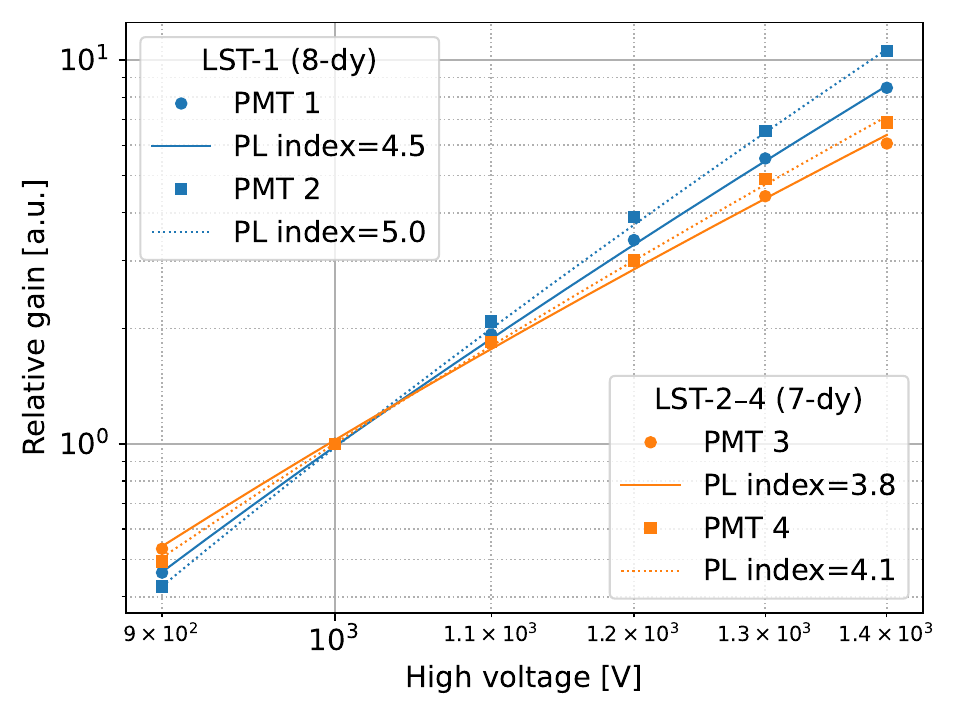}
    \caption{Examples of relative gain values measured against a HV ranging from 900 V to 1400 V and power-law (PL) curves fitted for them. The high-gain output of two PMTs for each of the LST-1 and LST-2--4 is plotted. Note that the incident light flux is not equal for these PMTs. The power law slope indices are shown in the legends.}  
    \label{fig:PMT-gaincurve-examples} 
\end{figure}

The PMT gain increases with higher HV, following a power-law curve, and the power-law slope varies for different tubes. 
Using the slope value, 
one can tune the gain for telescope observations.
In the module QC, we flashed the PMTs with the same light intensity, applying different HVs ranging from 900 V to 1400 V. Our interest lay in understanding the relative gain dependence on the HV for each PMT, and therefore a precise uniformity of the light is not required. The output charge against the applied HV was fitted by a power-law function:
$a(V-350)^{p}$, where $V$ is the HV, $a$ is the normalization and $p$ is the gain slope. The subtracted 350 V is the first dynode voltage with respect to the photocathode, which is fixed by the Zener diode as described in Section~\ref{ssec:PMT-structure}. In Figure~\ref{fig:PMT-gaincurve-examples}, one can find some examples of the relative gain values and power-law fitting.

\begin{figure}[!htb]
    \centering
    \includegraphics[width=0.45\textwidth]{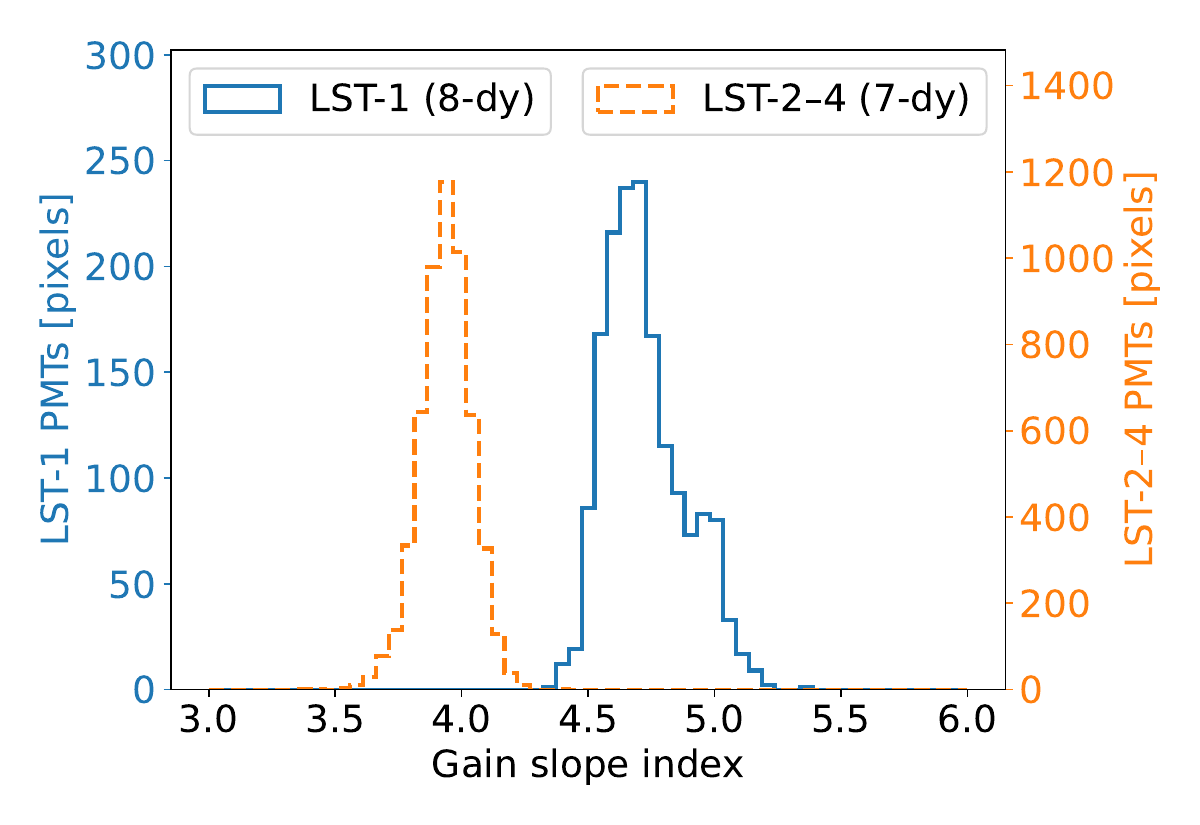}
    \caption{Distribution of the power-law fit slope of the gain-vs.-voltage curve. The solid and dashed lines represent the LST-1 and LST-2--4 modules, respectively. The scales for the LST-1 and LST-2--4 modules are shown on the left and right side of the figure, respectively.}   
    \label{fig:PMT-gainslope}
\end{figure}

Figure~\ref{fig:PMT-gainslope} depicts the distribution of measured slopes while Table~\ref{table:GainSlope} provides the 5\%, 50\%, and 95\% quantiles.
The LST-2--4 PMTs exhibit shallower slopes, likely due to their reduced dynode stages. On the other hand, the LST-1 modules show a wider spread and two-peak structure, resulting from a midway alteration in the production process of the dynodes to reduce the gain.

\begin{table}[hbtp]
  \caption{Quantiles of the power-law fit slope of the gain-vs.-voltage curve}
  \label{table:GainSlope}
  \centering
  \begin{tabular}{l|ccc}
    \multicolumn{4}{c}{Gain slope} \\
    \hline
    Quantiles & 5\%  & 50\% &  95\%  \\
    \hline
    LST-1  & 4.5  & 4.7 & 5.0 \\
    LST-2--4  & 3.8 & 3.9 & 4.1 \\
\end{tabular}
\end{table}

\subsubsection{Operation voltage}
\label{sssec:OperationVoltage}

The operation voltage is defined as the HV that yields a single p.e.\,charge of 93.7 ADC. This corresponds to an output pulse amplitude after the attenuation (see Section~\ref{ssec:pmt-performance}) equivalent to a PMT gain of 40000. It was determined for each tube to measure the HV-dependent characteristics in the QC. This value is calculated once the absolute gain at a certain HV is derived from the single-p.e.\,response, in addition to the gain-vs.-voltage slope\footnote{Due to the degradation of the PMTs, the operation voltage needs to be updated with an \textit{in situ} calibration method after they are installed in the telescope.}.

To simplify the process of measuring the single-p.e.\,response,
a different method was employed compared to the subtraction method described in Section~\ref{ssec:pmt-performance}. This was necessary due to the large amount of PMT data that needed to be analyzed within a limited time.
In this method, the output charge histogram is fitted with three Gaussian functions representing the 0-, 1-, and 2-p.e.\,components. Here, the small-charge tail of the single p.e.\, distribution (see 0--0.5 p.e.\,in Figure~\ref{fig:modeled_sphe-spectrum}) is ignored, and the gain is calculated from the mean of the 1-p.e.\,Gaussian function. 
This approach introduces a systematic bias of $\sim -20$ V in the operation voltage, which enlarges the pulse width by a few 
tens of ps. These systematics are below the required sensitivity of the QC, and they do not affect the results.

\begin{figure}[!htb]
    \centering
    \resizebox{0.45\textwidth}{!}{
    \includegraphics[width=\textwidth]{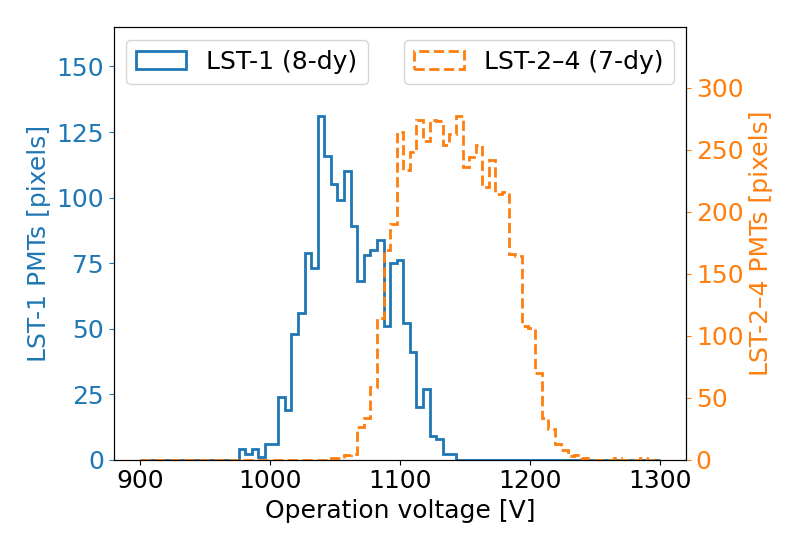}
    }
    \caption{Distribution of the PMT operation voltage. The solid and dashed lines represent the LST-1 and LST-2--4 modules, respectively. The scale for each of them is shown in the same way as in Figure~\ref{fig:PMT-gainslope}.}   
    \label{fig:PMT-Vop}
\end{figure}

Figure~\ref{fig:PMT-Vop} illustrates the distribution of the operation voltage for LST-1 and LST-2--4 PMTs. Table~\ref{table:Vop} provides the 5\%, 50\%, and 95\% quantiles for each distribution. Notably, the LST-2-4 PMTs require a higher voltage, $\sim 80$ V more, due to their reduced number of dynode stages.

\begin{table}[hbtp]
  \caption{Quantiles of the PMT operation voltage [V]}
  \label{table:Vop}
  \centering
  \begin{tabular}{l|ccc}
    Quantiles & 5\%  & 50\% &  95\%  \\
    \hline
    LST-1  & 1019  & 1059 & 1111 \\
    LST-2--4  & 1088 & 1140 & 1198 \\
\end{tabular}
\end{table}

\subsubsection{Pulse width}
To measure the pulse width of the output, we used the LD light of ${\sim} 4$ p.e./pulse as an input to the PMT modules. It should be noted that, contrary to the measurement described in Section~\ref{ssec:pmt-performance}, we irradiated an extended region of the photocathode determined by the LG exit, which may lead to the spread of the transit time of the photoelectrons from the photocathode to the first dynode ($\sim$ 300 ps in standard deviation). Also, the light pulse has a width of 800–920 ps in FWHM. We did not correct the contribution of these effects to the measured pulse width.

We obtained the pulse FWHM value by averaging the waveform from the high-gain channel of each PMT across thousands of pulses. To mitigate timing jitter, 
we aligned the centre of gravity of each pulses before superposing them.  We evaluated the FWHM by fitting the superposed waveform with a Gaussian function.

\begin{figure}[!htb]
    \centering
    \resizebox{0.45\textwidth}{!}{
    \includegraphics[width=\textwidth]{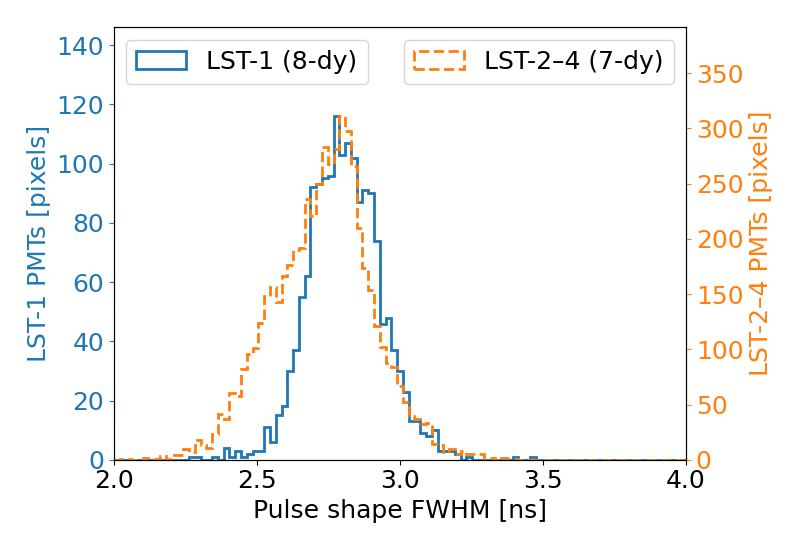}
    }
    \caption{Distribution of the PMT pulse width (FWHM). The solid and dashed lines represent the LST-1 and LST-2--4 modules, respectively. The scale for each of them is shown in the same way as in Figure~\ref{fig:PMT-gainslope}.} 
    \label{fig:PMT-pulsewidth}
\end{figure}

The distribution of the FWHM of the pulse time profile is shown in Figure~\ref{fig:PMT-pulsewidth}. The quantiles are in Table~\ref{table:PulseWidth}.
Those results satisfy the pulse width requirement of being shorter than 3 ns on average, and all tubes have a width lower than 3.5 ns.
These numbers are larger than those mentioned in Section~\ref{ssec:pmt-performance} because we illuminated an extended area of the PMT photocathode 
causing a spread in transit times of photoelectrons depending on the generated positions.

\begin{table}[hbtp]
  \caption{Quantiles of the PMT pulse width (FWHM)}
  \label{table:PulseWidth}
  \centering
  \begin{tabular}{l|ccc}
    \multicolumn{4}{c}{Pulse width [ns]} \\
    \hline
    Quantiles & 5\%  & 50\% &  95\%  \\
    \hline
    LST-1  & 2.6 & 2.8 & 3.0 \\
    LST-2--4  & 2.4 & 2.7 & 3.0 \\
\end{tabular}
\end{table}

\subsubsection{Signal-to-noise ratio}
The ratio of the 1-p.e.\,signal to the noise ($S/N$) is an important parameter for a precise measurement of faint Cherenkov signals, and
we derived it as follows.

The operation voltage was applied to each PMTs. 
The noise $N$ was quantified as the standard deviation of the extracted charge in a 5-ns integration window in the data without any light illumination.
The signal $S$ was assumed to be 93.7 ADC, which can be derived from the the elementary charge, the postulated gain of the PMT, the attenuation factor, the PACTA high-gain transimpedance, the amplification factor on the readout board and ADC conversion factor (200/pC). Though the actual gains might be slightly different from those figures, we used them because what we control during real observations is the signal amplitude including these factors.

\begin{figure}[!htb]
    \centering
    \resizebox{0.45\textwidth}{!}{
    \includegraphics[width=\textwidth]{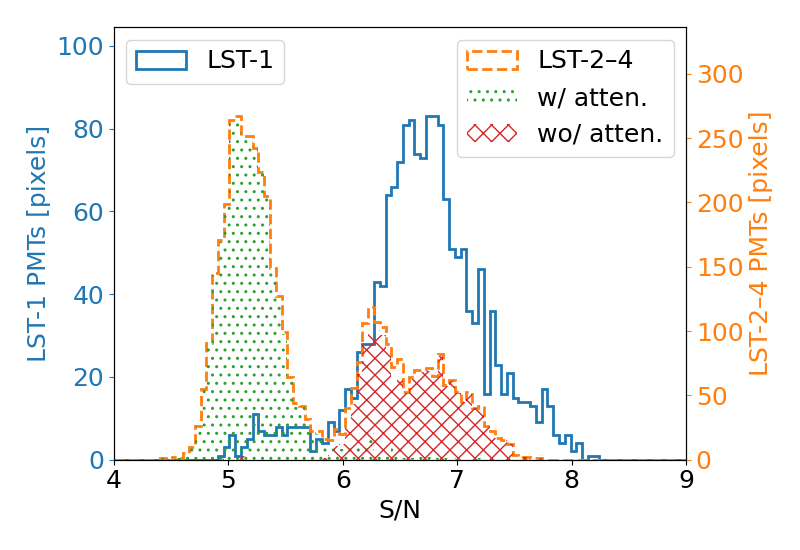}
    }
    \caption{Distribution of $S/N$ of the PMT modules. The blue solid and orange dashed lines represent the LST-1 and LST-2--4 modules, respectively. The scale for each of them is shown in the same way as in Figure~\ref{fig:PMT-gainslope}. The green-dot and red-criss-cross hatched histograms illustrate the variation in $S/N$ for subcategories of the LST-2--4 modules with and without PACTA attenuation, respectively.}   
    \label{fig:PMT-SN}
\end{figure}

The $S/N$ distributions are plotted in Figure~\ref{fig:PMT-SN}. The quantiles are presented in Table~\ref{table:PMT-SN}. 
The distributions show a bimodal population and
a considerable fraction have a lower $S/N$ for the LST-2--4 pixels.
We found that the signal from these pixels is attenuated at the PACTA board, as described in Section~\ref{sec:pacta}. Most pixels in the LST-1 modules also contain such an attenuation circuit, but the degradation of $S/N$ was not as pronounced as LST-2--4.
This difference was introduced due to a small change implemented in the electrical circuit on the PACTA board between LST-1 and LST-2--4.
However, the high $S/N$ of both LST-1 and LST-2--4 pixels enables us to characterize a shower image well, resolving a small charge at a ${\sim} 0.25$ p.e.\,level. This number is sufficiently small compared to other factors affecting the waveform amplitude, such as the Poisson fluctuations.
Any PMT with $S/N$ less than 4 was not found.
\begin{table}[hbtp]
  \caption{Quantiles of S/N of the PMT modules}
  \label{table:PMT-SN}
  \centering
  \begin{tabular}{l|ccc}
    \multicolumn{4}{c}{Signal-to-noise ratio} \\
    \hline
    Quantiles & 5\%  & 50\% &  95\%  \\
    \hline
    LST-1  & 5.7 & 6.2 & 6.8 \\
    LST-2--4 & 4.9 & 5.4 & 7.1 \\  
\end{tabular}
\end{table}

\subsubsection{Afterpulsing}
\label{ssec:afterpulsing}
In order to characterize PMT afterpulsing, PMTs were illuminated with a light pulse ($>100$ p.e./pulse) 50000 times and subsequent afterpulses were counted. The afterpulsing rate was derived by dividing the count number by the total photoelectron content of the input light.
This exposed the PMTs to $>5\times 10^6$ photoelectrons in total and yields only a statistical error of $\lesssim 3$\% for the specification afterpulsing rate of $2\times 10^{-4}$. While the afterpulsing occurs within a few microseconds, our data acquisition can cover only 1 \textmu s in one loop of DRS4 at once. To address this issue, we shifted the data-acquisition interval to cover two consecutive time ranges. Here, we present the results measured for up to 2 \textmu s after the light pulse. 
The systematic bias due to missing afterpulses occurring later than $2~\mu s$ is estimated to be ${\sim} 10$\%.

\begin{figure}[!htb]
    \centering
    \resizebox{0.45\textwidth}{!}{
    \includegraphics[width=\textwidth]{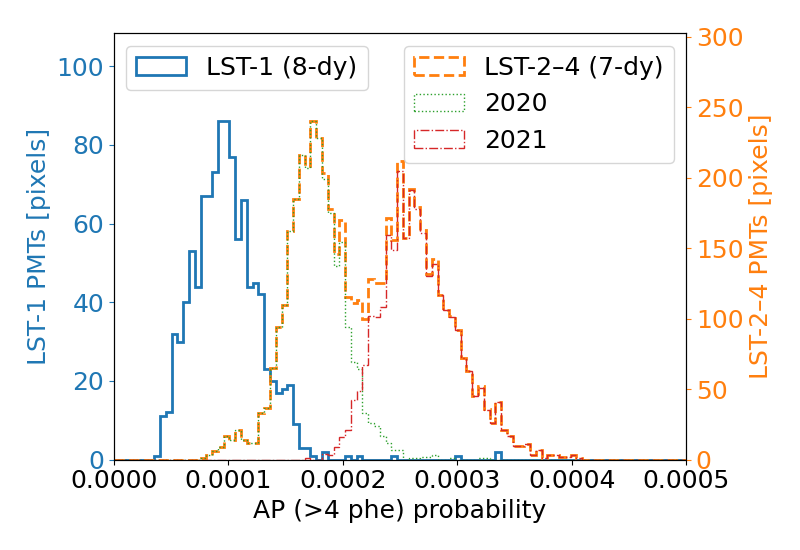}
    }
    \caption{Distribution of the PMT afterpulse (AP; $\geq 4$ p.e.) rate. The blue solid and orange dashed lines represent the LST-1 and LST-2--4 modules, respectively. The scale for each of them is shown in the same way as in Figure~\ref{fig:PMT-gainslope}. The green dotted and red dot-dashed line illustrates subcategories of the LST-2--4 modules measured in 2020 and 2021, respectively.}
    \label{fig:PMT-afterpulse}
\end{figure}

Figure~\ref{fig:PMT-afterpulse} displays the distribution of the measured afterpulsing rate above the threshold of 4 p.e., and Table~\ref{table:Afterpulse} summarizes the quantiles.
The values for LST-2--4 are higher than those of LST-1, possibly due to a longer time interval between production and measurement for the LST-2--4 PMTs, allowing more helium molecules in the atmosphere to penetrate the tubes\footnote{The R11920-100 tubes were delivered to us from 2012 to 2013, and they were measured about two years after that. The R12992-100 tubes were delivered from 2016 to 2018 and measured about four years later.}.
Furthermore, the distribution exhibits two peaks because the LST-2--4 QC was conducted during two different periods separated by nineteen months, suggesting that the afterpulse rate increased during those periods as mentioned.
The effect of this break is clearly seen in a system stability check with a reference PMT module (Figure~\ref{fig:Afterpulse_ref} in~\ref{sec:QC-stability}), and a similar increase has been found in our laboratory measurement. 
Despite this effect, the afterpulsing rate of most PMTs was $<4 \times 10^{-4}$ in this QC measurement.
We rejected only several PMTs whose afterpulsing rate exceeded this limit.
A long-term evolution of the afterpulsing rate after the telescope operation started showed that it does not keep increasing. The details will be discussed in another paper.

\begin{table}[hbtp]
  \caption{Quantiles of the afterpuling rate of each PMT}
  \label{table:Afterpulse}
  \centering
  \begin{tabular}{l|ccc}
    \multicolumn{4}{c}{Afterpulsing rate [$\times 10^{-4}$]} \\
    \hline
    Quantiles & 5\%  & 50\% &  95\%  \\
    \hline
    LST-1  & 0.6 & 1.0 & 1.5 \\
    LST-2--4 & 1.4 & 2.2 & 3.2 \\    
\end{tabular}
\end{table}

\subsubsection{Linearity}
The linearity of the signal detection was tested using the fast LD pulser together with the two filter wheels mentioned in Section~\ref{sect:qcsetup}. 
The laser intensity was varied from ${\sim} 4$ to ${\sim} 1000$~p.e./pixel for LST-1 modules\footnote{It should be noted that the linearity of the readout electronic including the preamplifier was independently tested up to 4000 p.e.\,and verified up to 2000 p.e.\,using the test pulse injection from the SCB for all LST-1 and LST-2--4 modules.} and from ${\sim} 4$ to ${\sim} 3000$~p.e./pixel for LST-2--4 modules.
Here, the input light amount is converted to the photoelectron number as follows. First, we estimated the incident photoelectron number for one specific filter combination, with which the PMT signal was neither affected by saturation or noise, using the gain obtained in Section~\ref{sssec:OperationVoltage}. Next, we derived the number for the other combinations by dividing this figure by the relative opacity explained in Section~\ref{sect:qcsetup}. We then fit the output charge against the relative opacity by a linear function for each PMT-module channel.

\begin{figure}[ht]
\begin{center}
\resizebox{0.45\textwidth}{!}{
    \includegraphics[width=\textwidth]{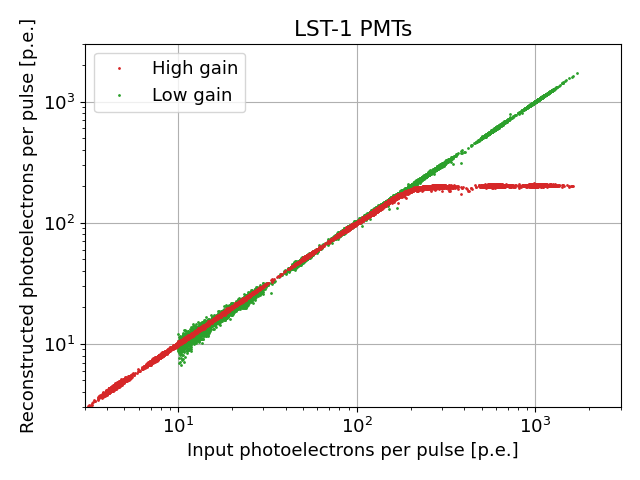}
    }
\resizebox{0.45\textwidth}{!}{
    \includegraphics[width=\textwidth]{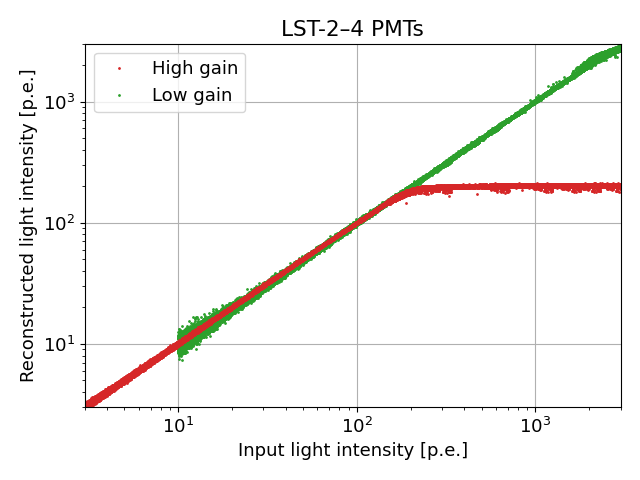}
    }
\caption{Output of the PMT modules in photoelectrons against the input pulse light incident on the PMTs per pulse in photoelectrons. The output charge is converted to the reconstructed photoelectron number. Each marker corresponds to one measurement of each channel under each light intensity. Red squares and green circles represent the high and low gain channels, respectively. {\it top:} LST-1 PMTs and {\it bottom:} LST-2--4 PMTs.}
 \label{fig:Output-Input}
\end{center}
\end{figure}

\begin{figure}[ht]
\begin{center}
\resizebox{0.45\textwidth}{!}{
    \includegraphics[width=\textwidth]{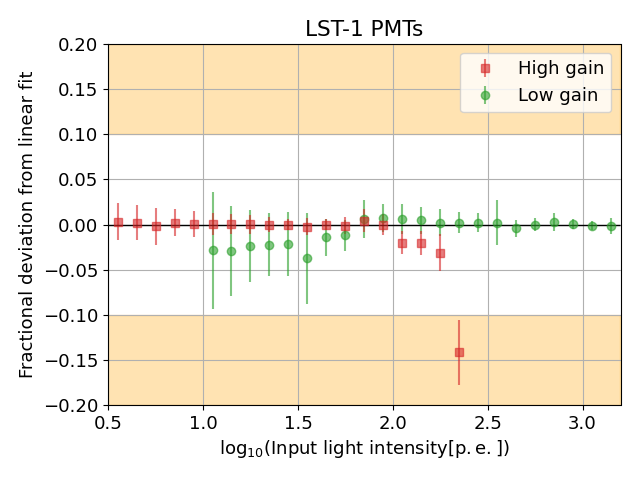}
    }
\resizebox{0.45\textwidth}{!}{
    \includegraphics[width=\textwidth]{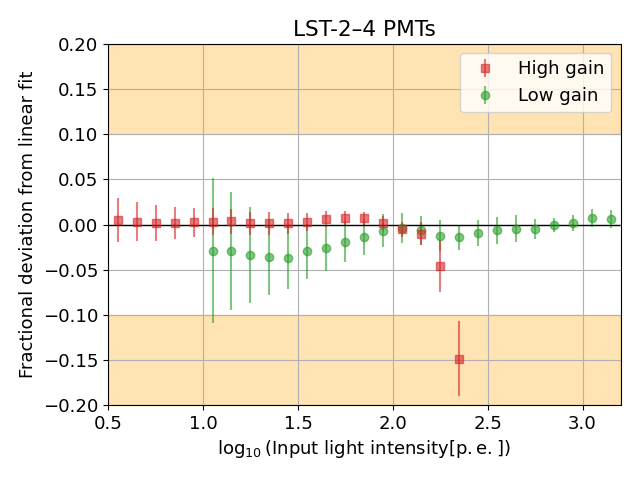}
    }    
\caption{Linearity of the PMT modules. The ordinate represents distributions of the fractional deviation from a linear fit of the output response of each channel. The abscissa represents the pulse light incident on the PMTs in photoelectrons. Each marker and error bar represent the mean and the standard deviation, respectively, in each intensity bin. Red squares and green circles represent the high and low gain channels, respectively. {\it top:} LST-1 PMTs and {\it bottom:} LST-2--4 PMTs.}
 \label{fig:Linearity}
\end{center}
\end{figure}

The output of each channel for each incident photoelectron number is plotted in Figure~\ref{fig:Output-Input}. On the one hand, the high-gain output follows the linearity from $\sim 4$ p.e.\,to $\sim 200$ p.e., and saturates above the region. On the other hand, the low gain keeps the linearity even around 1000 p.e.\,although it is affected by noise below $\sim$ 30 p.e..
To compare the deviations with our criterion, the results are illustrated in Figure~\ref{fig:Linearity} as relative differences from the linear relation.
By choosing one of the high or low gain outputs depending on the light intensity, most of the pixels fulfilled the requirement of the linearity between 4 and 2000 p.e.\,to be within 10\%, and the others were rejected. This allows the cameras to satisfy our requirement for the fractional charge resolution of $<10$\% at 1000~p.e.~\cite{CTAReq}.

\subsubsection{Crosstalk}
\begin{figure}[!htb]
    \centering
    \resizebox{0.5\textwidth}{!}{
    \includegraphics[width=15cm]{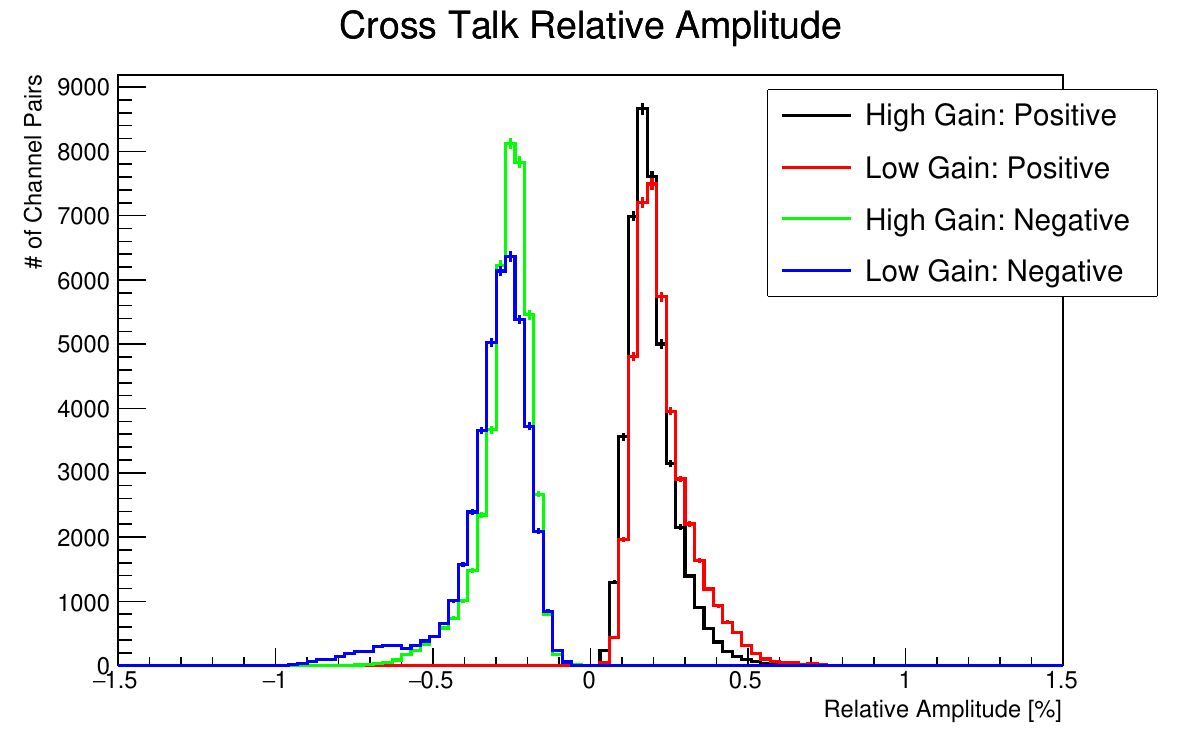}
    }
    \caption{Distribution of relative cross talk amplitude for 265 boards. Both positive and negative crosstalks were examined in both high and low-gain channels within the same board. }
    \label{fig:CT}
\end{figure}

Crosstalks between readout channels have been evaluated using the test pulses from the SCB. Large pulses (${\sim} 70$ p.e. for the high gain and ${\sim} 2000$ p.e.\,for the low gain) were injected to one readout channel, and a signal peak was searched in the recorded waveforms of the surrounding six other channels in the same readout board. As crosstalk in electronics are generally electromagnetic interference, both positive and negative peaks could be seen in the recorded waveforms. Therefore, crosstalks in both polarities were checked. The ratio of the height of the found peak to that of the injected pulse was defined as relative crosstalk amplitude and plotted in Figure \ref{fig:CT}. We specified the tolerance of crosstalk (both polarities) as 1\% and none of the channels exceeded this limit as can be seen in the figure. 

\subsection{Qualification}
Summarizing, in the end of the QC process we selected only those pixels that satisfied the following criteria:
\begin{itemize}
    \item Pulse width $\leq \SI{3.5}{ns}$ in FWHM at the operatin voltage
    \item Single-p.e.\,signal-to-noise ratio $\geq 4$ at the operatin voltage
    \item Afterpulsing rate above 4 p.e.\footnote{We remark that the LST-1 afterpulsing measurement was done with the individual PMT QC.} $\leq 4 \times 10^{-4}$ 
    \item Deviation from the linearity between 4 and ${\sim} 2000$ p.e. $\leq$ 10\% 
    \item Crosstallk with test pulses $\leq 1\%$ in the pulse height.
\end{itemize}
Of the 2019 PMTs checked for the LST-1 telescope, 2002 successfully passed the tests. We then used the qualified PMTs to build 265 PMT modules for the first telescope and additional six as spares. For LST-2--4, we repaired or replaced the disqualified components and subjected them to another round of qualification. 
Ultimately, 5652 out of 5695 PMTs successfully passed the tests, and 5565 PMTs were used for the 795 modules of LST-2-4.
 

\section{Conclusion}
We have developed the PMT modules for the cameras of the LSTs of the CTAO and characterized their performance. 
The modules achieved the target requirements, which will lead to the maximum performance of the telescopes.

 The mass production of all modules for the four LSTs of the northern CTAO has been completed.
 The quality control of them has been performed in the form of a 19-module mini-camera, which resulted in smooth and efficient measurements. Based on these results, less than 1\% of the PMTs were rejected. By replacing those disqualified PMTs and other erroneous electrical elements, we have got 271 and 795 qualified modules for LST-1 and LST-2--4 respectively, which are  to build four LST cameras. 

\section*{Author Contribution}
{\bf T. Saito:} project leadership, management of PMT module assembly and QC measurements, development of QC setup, paper-drafting, and edition; {\bf M. Takahashi:} project leadership, development of QC setup, PMT measurements for LST-1, paper drafting and edition; {\bf Y. Inome:} development of QC setup, management of PMT modules assembly;
{\bf M. Artero, O. Blanch and D. Kerszberg:} development and quality control of trigger mezzanine boards; {\bf S. Nozaki} precise analysis of LST-1 modules and development of readout boards; {\bf Y. Kobayashi:} precise measurements of PMT pulse shapes; {\bf Y. Konno, and H. Kubo:} development of readout boards; {\bf D. Mazin:} definition of QC and characterization tests and the development of control software; {\bf D. Nakajima:} management of PMT module assembly and QC measurements for LST-1 modules; {\bf H. Ohoka:} development of SCBs; {\bf A. Okumura, T. Yamamoto}, and {\bf T. Yoshida:} development and measurements of light guides; {\bf S. Sakurai:} Outstanding contribution to module assembly and QC {\bf M. Teshima:} development of PMTs with Hamamatsu photonics and funding acquisition. The rest of the authors have contributed to QC measurements or/and debugging of module elements.

\section*{Acknowledgements:}
We gratefully acknowledge financial support from the following agencies and organisations: ICRR, University of Tokyo, JSPS, MEXT, Japan; IFNN, Italy; the Spanish Ministry of Science and Innovation and the Spanish Research State Agency through the grant  "FPN: PID2019-107847RB-C41”, the “CERCA'' program funded by the Generalitat de Catalunya.
We also thank the Instituto de Astrofísica de Canarias (IAC) for the technical support and for the use of the clean room facilities of the IAC, in the central headquarters in La Laguna and of IACTEC.
This research is part of the Project RYC2021-032991-I, funded by MICIN/AEI/10.13039/501100011033, and the European Union “NextGenerationEU”/RTRP.

\appendix
\section{Setup for the PMT QC for LST-1}
\label{sec:PMTQC}
For LST-1, we first performed a QC of individual elements alone, namely, PMT units, SCBs, and readout boards.
We screened malfunctioning products and assembled the components into modules. Also, we performed a measurement to qualify the afterpulsing rate of the PMTs, which was not done in the assembled module QC.
For LST-2--4, we did not perform the QC of PMTs alone before assembling modules, as the error rate was proven to be low during the LST-1 QC. The afterpulse measurement was done also in the module QC.

As a light source, we used the same laser diode and driver circuit described in Subsection~\ref{sect:qcsetup}.
The light was fed into the dark box by an optical fibre, and the intensity was controllable within a three-orders-of-magnitude range using two optical filter wheels. 
We measured the relative opacity of these filter combinations by illuminating several LST PMTs with the constant-intensity LD and taking the average output of the PMTs.
As the PMT QC measurements were performed before the completion of readout boards, the output signals from the PMT unit were recorded by DRS4 evaluation boards~\footnote{\url{https://www.psi.ch/en/drs/evaluation-board}}.

The sampling speed was set between 1--5 GHz depending on the measured characteristics. In Section~\ref{ssec:afterpulsing}, we report the result of the PMT afterpulsing, which was measured with a 1 GHz sampling rate.
The data acquisition was initiated by a trigger signal synchronized with the LD pulser.
The PMT HV was configured using a simple custom-built circuit. 
It sets the HV a required value between 0 and 1.5 kV via a single-board computer unit (Raspberry Pi Model B with 512MB RAM) and a DAC (Texas Instruments DAC7578).
The HV output was regulated by a reference (Linear Technology LTC6652) with an accuracy of $\pm 0.5$\%.
This system could host and measure four PMTs (eight PMTs after an upgrade) simultaneously.

\section{Stability of the module QC system}
\label{sec:QC-stability}
During the measurements of the LST-2--4 modules, we kept one module in the center of the module rack as a reference detector. In this section, we report the stability of the system based on the results of this module.

\begin{figure}[!htb]
    \centering
    \resizebox{0.45\textwidth}{!}{
    \includegraphics[width=\textwidth]{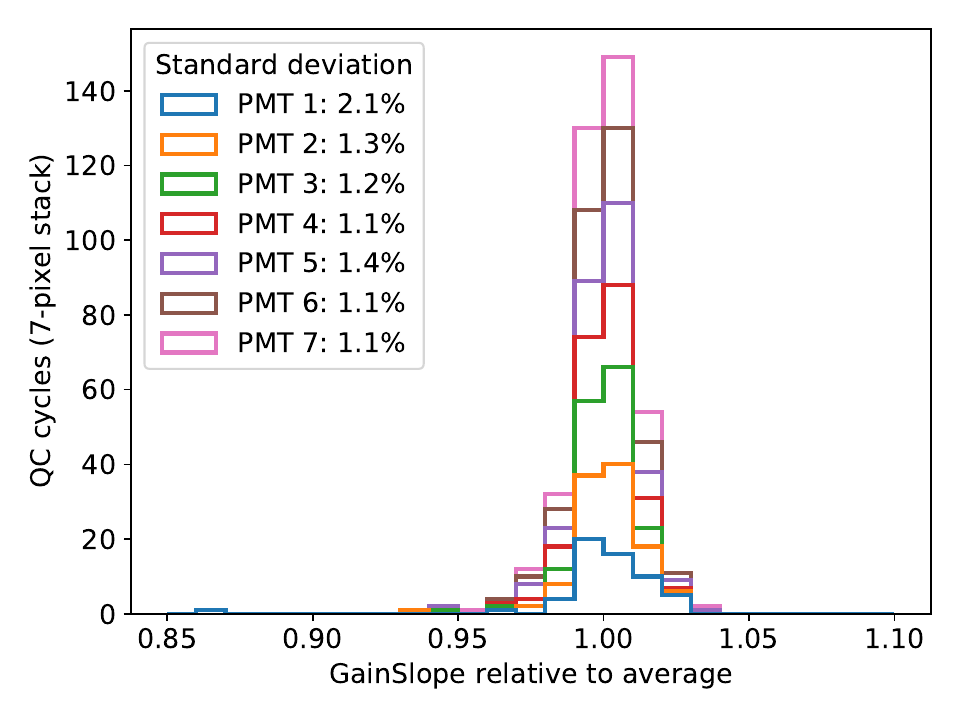}
    }
    \caption{Distribution of the gain slope of the reference PMT modules.}   
    \label{fig:GainSlope_ref}
\end{figure}

\paragraph{Gain-vs.-voltage slope}
Figure~\ref{fig:GainSlope_ref} depicts a spread of the gain slope value of seven PMTs of the reference module measured in every QC measurement cycle. The standard deviation ranges from $\sim 1$\% to $\sim 2$\%. 
Such variations are negligible for the gain control as other factors,  e.g.\,variation from the aging of the individual PMT gains, are more dominant.

\begin{figure}[!htb]
    \centering
    \resizebox{0.45\textwidth}{!}{
    \includegraphics[width=\textwidth]{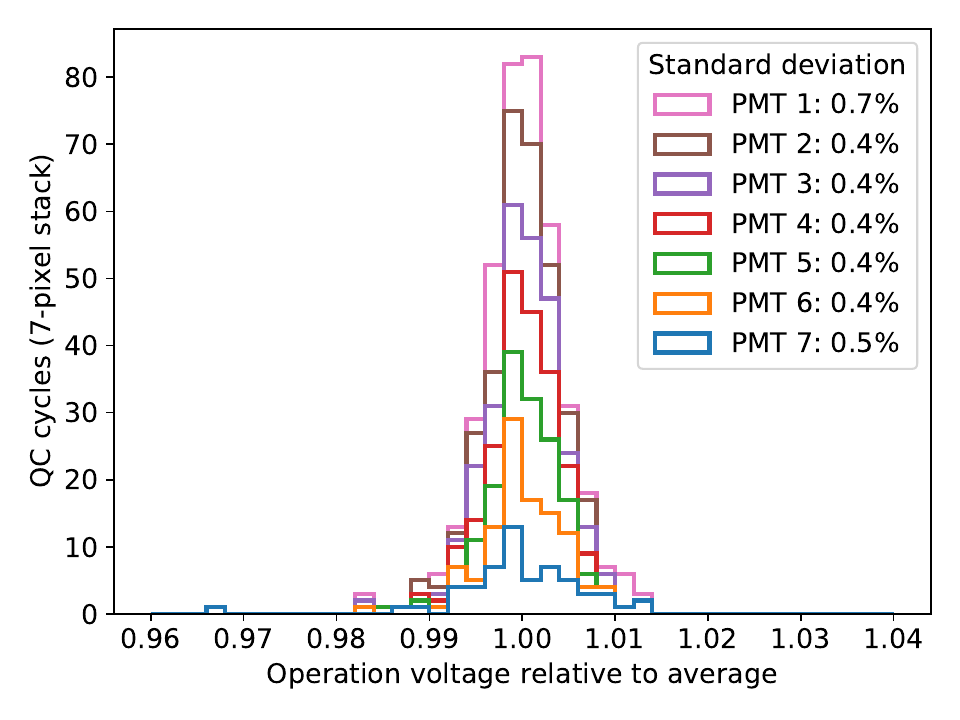}
    }
    \caption{Distribution of the operation voltage of the reference PMT modules.}   
    \label{fig:Vop_ref}
\end{figure}

\paragraph{Operation voltage}
The standard deviation of the operation voltage ranges from 0.4\% to 0.7\% as shown in Figure~\ref{fig:Vop_ref}. These numbers are also negligible from a point of view of configuring the PMT modules as in real observations.

\begin{figure}[!htb]
    \centering
    \resizebox{0.45\textwidth}{!}{
    \includegraphics[width=\textwidth]{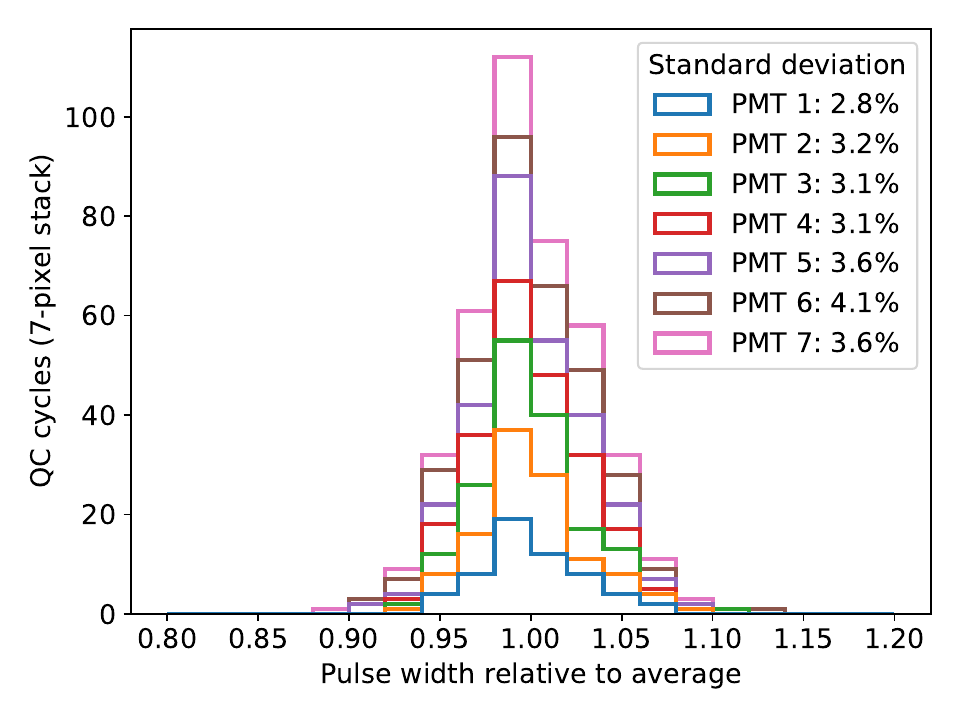}
    }
    \caption{Distribution of the pulse width (FWHM) of the reference PMT modules.}   
    \label{fig:PulseWidth_ref}
\end{figure}

\paragraph{Pulse width}
The standard deviation of the pulse width ranges from $\sim 3$\% to $\sim 4$\% as shown in Figure~\ref{fig:PulseWidth_ref}. It is sufficiently small compared to other factors involved in this measurement, for example, the LD light pulse width.

\begin{figure}[!htb]
    \centering
    \resizebox{0.45\textwidth}{!}{
    \includegraphics[width=\textwidth]{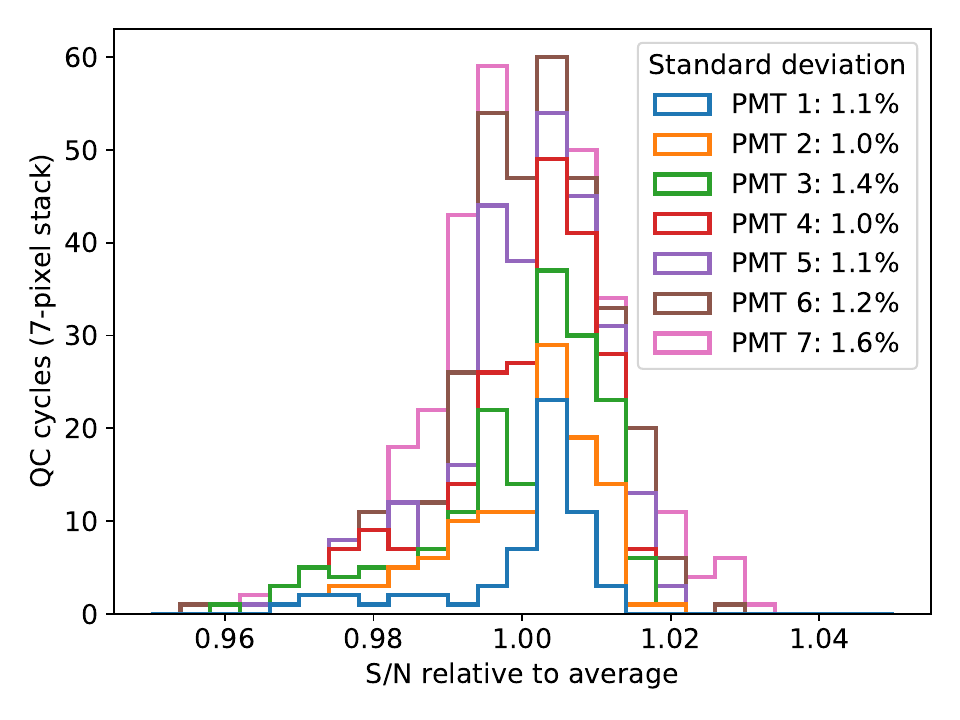}
    }
    \caption{Distribution of $S/N$ of the reference PMT modules.}   
    \label{fig:SN_ref}
\end{figure}

\paragraph{Signal-to-noise ratio}
The standard deviation of the pulse width ranges from $\sim 1$\% to $\sim 2$\% as shown in Figure~\ref{fig:SN_ref}. Such an error is negligible for screening PMT modules with $S/N<4$.

\begin{figure}[!htb]
    \centering
    \includegraphics[width=0.45\textwidth]{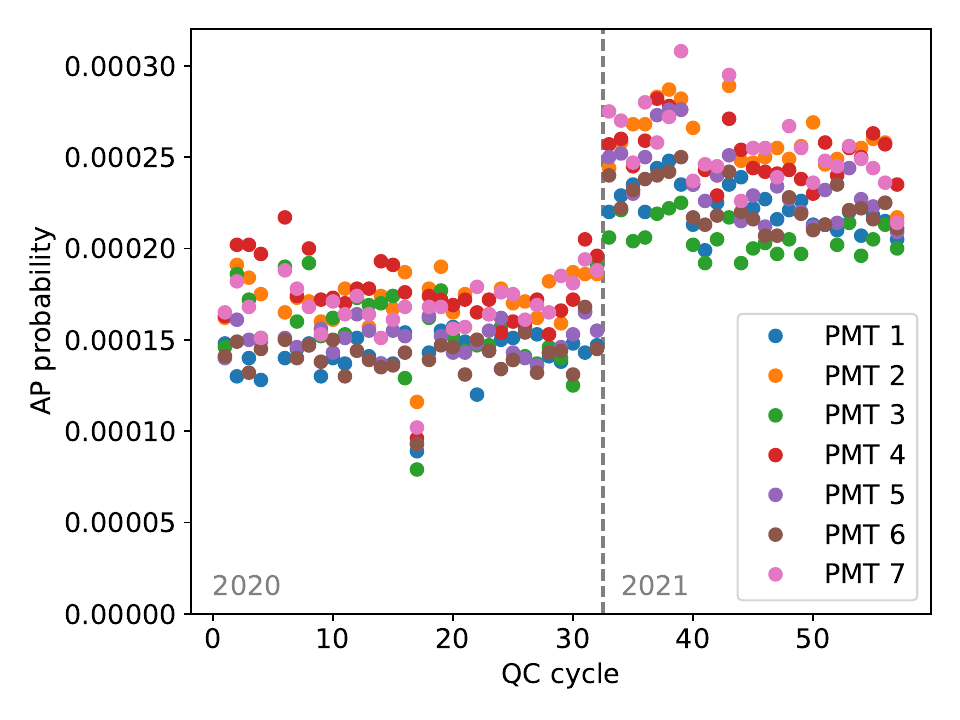}
    \caption{Measured value of the afterpulsing (AP) probability of seven PMTs in the reference modules for each QC cycle.}
    \label{fig:Afterpulse_ref}  
    \end{figure}

\begin{figure}[!htb]
    \centering
          \subfigure{
        \includegraphics[width=0.45\textwidth]{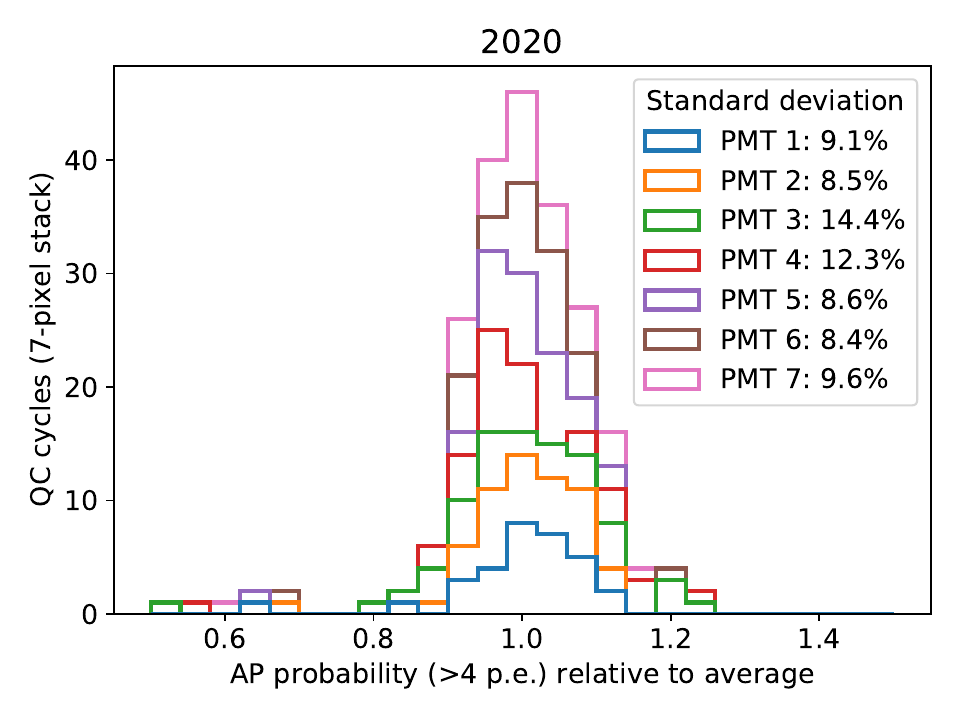}
        \label{fig:Afterpulse2020_ref}
    }
     \subfigure{
        \includegraphics[width=0.45\textwidth]{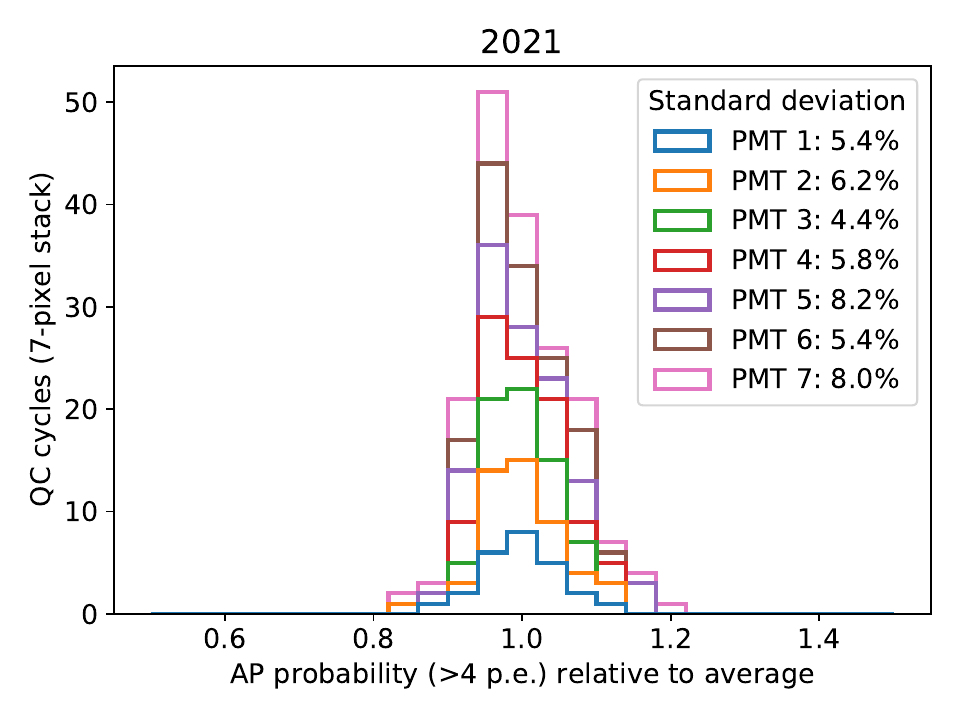}
        \label{fig:Afterpulse2021_ref}
    }    
    \caption{Distribution of the AP probability of the reference PMTs for QC cycles in 2020 (top) and 2021 (bottom)}   
\end{figure}

\paragraph{Afterpulsing}
As mentioned in Section~\ref{ssec:afterpulsing}, the afterpulsing probability exhibit a large gap between the measurements in 2020 and in 2021, which are separated by a nineteen-month break. This is also clearly demonstrated in Figure~\ref{fig:Afterpulse_ref}. In each of these periods, the standard deviation ranges from $\sim 5$\% to $\sim 14$\% (Figure~\ref{fig:Afterpulse2020_ref},~\ref{fig:Afterpulse2021_ref}). Since the afterpulsing probability changes by a factor of two, such a measurement error is acceptable.

\begin{figure}[!htb]
    \centering
    \resizebox{0.45\textwidth}{!}{
    \includegraphics[width=\textwidth]{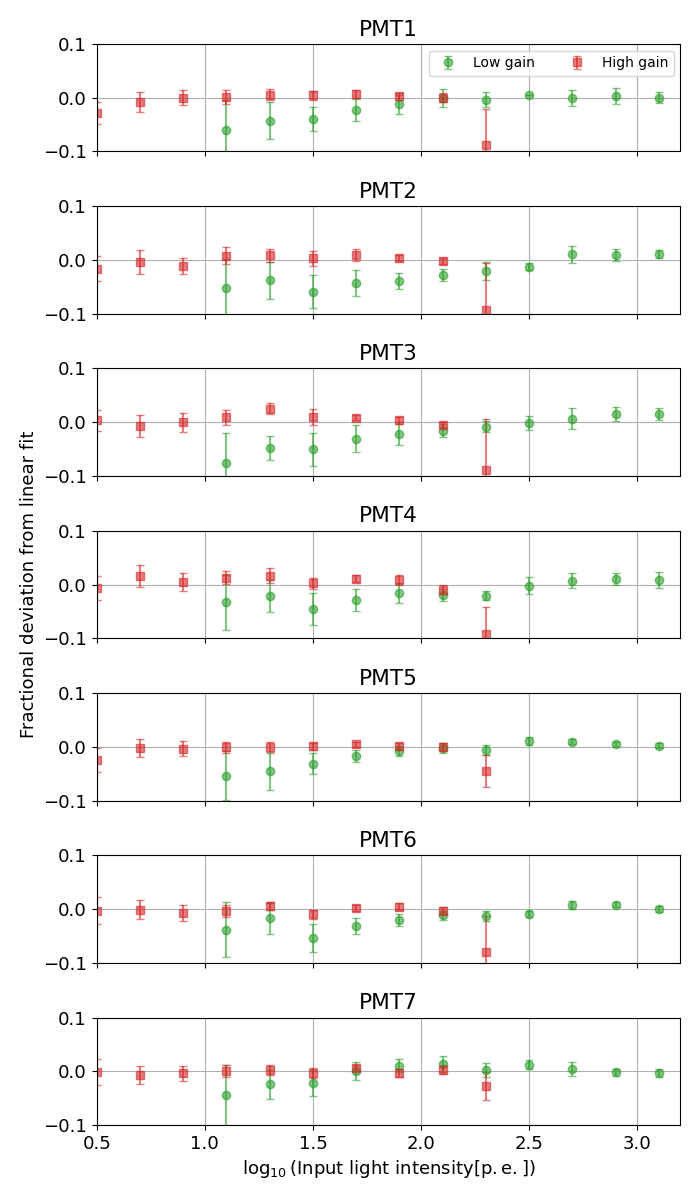}
    }
    \caption{Linearity of the reference PMT module. Red squares and green circles represent the high- and low-gain channels, respectively. }   
    \label{fig:Linearity_ref}
\end{figure}

\paragraph{Linearity}
In Figure~\ref{fig:Linearity_ref}, the fractional deviation from the linearity of the reference PMT module is illustrated. The standard deviation in each bin in the high- and low-gain linear regions ranges from $<1$\% to $\sim 3$\%, which is much lower than our criterion for the fractional deviation, 10\%.

\bibliography{mybibfile}

\end{document}